\documentclass[10pt,journal,compsoc]{IEEEtran}
\usepackage{multirow}
\usepackage{amsthm,amsmath}
\usepackage{caption}
\usepackage{xcolor}
\usepackage{subfigure}
\usepackage[ruled,linesnumbered]{algorithm2e}
\usepackage{enumitem}
\usepackage[most]{tcolorbox}
\usepackage{balance}
\usepackage{url}
\usepackage{booktabs}
\usepackage{amssymb}
\usepackage{cellspace}
\usepackage{colortbl}
\usepackage{array,multirow}
\usepackage{siunitx}
\usepackage[colorlinks=true]{hyperref}
\usepackage{pifont}
\usepackage{amsmath,amssymb,amsfonts}
\usepackage[ruled,linesnumbered]{algorithm2e}
\usepackage{algorithmic}
\algsetup{linenosize=\small}
\SetKwInput{KwInput}{Input}                %
\SetKwInput{KwOutput}{Output}              %

\usepackage{xcolor}
\usepackage{enumitem}
\usepackage{amsmath,amssymb}

\ifCLASSOPTIONcompsoc
  \usepackage[nocompress]{cite}
\else
  \usepackage{cite}
\fi

\ifCLASSINFOpdf
\else
\fi

\begin{document}
\newcommand{\toolname}{{\sc Gotcha}\xspace}

\newcommand{\blue}[1]{\textcolor{cyan}{#1}}
\newcommand{\yz}[1]{\mynote{Zhou}{#1}}
\title{Gotcha! This Model Uses My Code! Evaluating Membership Leakage Risks in Code Models}

\author{Zhou~Yang, Zhipeng Zhao, Chenyu Wang, Jieke Shi, Dongsun Kim, DongGyun Han and~David~Lo~\IEEEmembership{Fellow,~IEEE}%
\IEEEcompsocitemizethanks{
\IEEEcompsocthanksitem Z. Yang, C. Wang, J. Shi, D. Lo are
with the School of Computing and Information Systems, Singapore
Management University. \protect\\
E-mail: \{zyang, chenyuwang, jiekeshi, davidlo\}@smu.edu.sg
\IEEEcompsocthanksitem Z. Zhao is with University of Copenhagen. E-mail: zhipeng.zhao@di.ku.dk.
\IEEEcompsocthanksitem D. Kim is with Korea University, South Korea. E-mail: darkrsw@gmail.com.
\IEEEcompsocthanksitem D. Han is with Royal Holloway, University of London. E-mail: DongGyun.Han@rhul.ac.uk.
}%
}

\markboth{Journal of IEEE Transactions on Software Engineering,~Vol.~14, No.~8, Feb~2024}%
{Shell \MakeLowercase{\textit{et al.}}: Bare Demo of IEEEtran.cls for Computer Society Journals}
\IEEEtitleabstractindextext{%

\begin{abstract}
  Leveraging large-scale datasets from open-source projects and advances in large language models, recent progress has led to sophisticated code models for key software engineering tasks, such as program repair and code completion. These models are trained on data from various sources, including public open-source projects like GitHub and private, confidential code from companies, raising significant privacy concerns.
  This paper investigates a crucial but unexplored question: \textit{What is the risk of membership information leakage in code models?} 
  Membership leakage refers to the vulnerability where an attacker can infer whether a specific data point was part of the training dataset. 
  We present \toolname, a novel membership inference attack method designed for code models, and evaluate its effectiveness on Java-based datasets.
  \toolname simultaneously considers three key factors: model input, model output, and ground truth. Our ablation study confirms that each factor significantly enhances attack performance.
  Our ablation study confirms that each factor significantly enhances attack performance.
  Our investigation reveals a troubling finding: \textbf{membership leakage risk is significantly elevated}.
  While previous methods had accuracy close to random guessing, \toolname achieves high precision, with a true positive rate of 0.95 and a low false positive rate of 0.10.
  We also demonstrate that the attacker's knowledge of the victim model (e.g., model architecture and pre-training data) affects attack success. Additionally, modifying decoding strategies can help reduce membership leakage risks.
  This research highlights the urgent need to better understand the privacy vulnerabilities of code models and develop strong countermeasures against these threats.

  \end{abstract}

\begin{IEEEkeywords}
  Membership Inference Attack, Privacy, Large Langauge Models for Code, Code Completion
\end{IEEEkeywords}}

\maketitle

\section{Introduction}
\label{sec:intro}

Recent years have witnessed a surging trend in training large language models~\cite{bert,RoBERTa,T5,plbart} on source code~\cite{CodeXGLUE,husain2019codesearchnet} to produce \textit{code models} for a wide range of critical software engineering tasks~\cite{codellm_survey}, including code completion~\cite{9794048}, software contents summarization~\cite{9825884,TechSumBot} and defect prediction~\cite{9462962}.
For example, GitHub Copilot uses the OpenAI Codex~\cite{codex} model that is trained on billions of lines of code to assist developers in writing code. Similar tools include IntelliCode,\footnote{\url{https://visualstudio.microsoft.com/services/intellicode/}} and CodeWhisperer,\footnote{\url{https://aws.amazon.com/codewhisperer/}} which have been integrated into popular IDEs such as Visual Studio Code.

Although code models have achieved noticeable success in both academic and
industrial settings, a series of studies expose that these models are
vulnerable to various attacks, including adversarial
attacks~\cite{alert,Yefet2020,Epresentation2021,9825895,nguyen2023adversarial}, data poisoning
attacks~\cite{263874,you-see,coffee}, and privacy leakage~\cite{yang2023memorzation,291327}.
The vulnerabilities may introduce new defects, which are often difficult to discover and fix. Thus, it is necessary to clearly assess the potential risks of using code models.

This study explores a critical, previously overlooked aspect of code models: their vulnerability to \textit{membership inference attacks} (MIA). 
MIA aims to determine whether a specific data instance was included in the training dataset of a \textit{victim model}—the model subjected to the attack.
Users and developers of code models can be interested in discerning such membership information for various purposes.
For example, users may avoid a model if they discover it was trained on low-quality data, such as code containing smells or non-standard practices~\cite{10006873}.
Membership information can also be used to detect unauthorized training, i.e., whether the model is trained on the code that is not authorized for training~\cite{CoProtector}, to protect the intellectual property of the code.
Attackers, however, can exploit MIA to uncover further vulnerabilities. For instance, recent research on backdoor attacks~\cite{you-see,advdoor} shows that if malicious code is part of the training data, the attacker could use MIA to identify this and trigger backdoor vulnerabilities to compromise the model.

Additionally, the severity of MIA is further magnified due to the diversity of the training data sources for code models.
These sources include public domains, like open-source projects on GitHub, as well as private repositories with confidential corporate code. For example, Amazon CodeWhisperer, an automatic code completion tool, is trained on both open-source and proprietary datasets~\cite{AWSCodeWhisperer}, which may include sensitive data such as API keys and personal information~\cite{basak2023secretbench}. Yang et al.~\cite{yang2023memorzation} show that code models can generate API keys, and attackers could use MIA to infer whether such keys are part of the training data, potentially exposing sensitive information from companies.

Growing concerns about code models drive us to explore an essential yet underexamined question: \textit{what is the risk of membership information leakage in code models?} 
To answer this, we introduce \toolname, a novel membership inference attack method for code completion models.
First, a \textit{surrogate model} is trained to mimic the victim model's behavior. 
Then, the surrogate is provided with inputs from both training and non-training data to generate outputs.
Using these outputs, we train a classifier to distinguish between training and non-training data.
This classifier encodes three types of information—model input, model output, and ground truth—into embeddings.
Our ablation study confirms that each type of information contributes to classifier performance.

Our research focuses on highlighting the membership leakage risks posed by \toolname.
As a demonstrative measure, our experimental framework is centered around CodeGPT~\cite{CodeXGLUE}, an open-source model that demonstrates competitive performance on the code completion task.
We additionally evaluate the generalizability of \toolname on five open-source models, including \texttt{CodeGen}~\cite{codegen}, \texttt{CodeParrot}~\cite{codeparrot}, \texttt{gpt-neo}~\cite{gpt-neo}, \texttt{PolyCoder-160M}, and \texttt{PolyCoder-0.4B}~\cite{10.1145/3520312.3534862}.
We fine-tune these models on the JavaCorpus dataset~\cite{javacorpus} to obtain the victim models.

We consider two baseline MIA methods.
The first baseline, proposed by Hisamoto et al.\cite{hisamoto-etal-2020-membership}, involves training classifiers such as nearest neighbors and decision trees. These classifiers utilize statistical features of model outputs as their inputs.
The second baseline\cite{carlini21extracting} adopts a ranking mechanism, utilizing language-centric metrics such as perplexity. In this approach, data instances at higher rankings are deemed more likely to be members of the training data.
Our experiments have been executed across a spectrum of configurations, which encompasses variations in hyper-parameters such as the number of training epochs, the selection of surrogate models, etc.

We conduct our experiment on a Java dataset.
The experiment results show that the proposed method achieves the best performance in terms of both the attacker's \textit{power} (i.e., the true
positive rate) and the attacker's \textit{error} (i.e., the false positive rate).
Utilizing CodeGPT as the surrogate model, \toolname demonstrates a power value of
$0.95$, substantially surpassing the two baseline methods, which approximate the randomness of guessing.
We unveil a concerning fact: \textbf{the risk associated with the leakage of membership information is elevated}.

Further exploration reveals that the extent of the attacker's knowledge of the victim model affects the membership information leakage risks.
To be more specific, the attacker can infer membership with a higher accuracy if the attacker can access a larger portion of the victim model's training data, suggesting that the model developers should include more training data
that is inaccessible to the attacker.
It also favors the attacker if the attacker uses a surrogate model that shares the same architecture as the victim model.
For example, using the CodeGPT as the surrogate model can achieve a higher power value than using the 12-layer Transformer or LSTM as the surrogate model.
As a result, to protect code models from MIA, the model developers should try to conceal the details of the victim model's architecture.

We further investigate how the decoding strategy affects the attacking results.
By default, CodeGPT uses beam-search~\cite{beam-search} to generate the outputs.
We find that using a different decoding strategy (i.e., top-$k$ sampling) can mitigate the risk of MIA.
This paper calls for attention to the privacy concerns on code models and developing effective defense strategies against such attacks.
The contributions of this paper include:
\begin{itemize}[leftmargin=*]
  \item {\bf MIA threats in code models:} We are the first to investigate the risks of membership information leakage when using code models. We propose \toolname, an effective membership inference attack method for code models to investigate such risks. We evaluate the proposed method on the CodeGPT model, demonstrating that there exists a high risk of membership information leakage.
  \item {\bf Risk assessment of code models:} The attacker's knowledge of the victim model affects the risk of membership information leakage. Knowing the victim model's architecture and accessing a larger portion of the training data can increase the risk. We also find that using a different decoding strategy (i.e., changing from beam-search to top-$k$ sampling) can mitigate the risk.
  \item {\bf Replication Package:} To facilitate further studies in evaluating the such risks and developing an effective defense against such attacks, we make our code and data publicly available at \url{https://github.com/yangzhou6666/MIA-LLM4Code}
\end{itemize}

\noindent \textbf{Paper Structure.} This paper unfolds in a structured manner as delineated hereafter.
Section~\ref{sec:background} describes the background of this study, including the code models and motivation for studying privacy attacks.
In Section~\ref{sec:methodology}, we explain our proposed method.
Section~\ref{sec:exp} presents the experiment settings.
Section~\ref{sec:result} evaluates the risks by answering research questions.
We further discuss potential defensive strategies and threats to validity in Section~\ref{sec:discussion}.
Section~\ref{sec:related_work} introduces relevant works.
Finally, we conclude the paper and provide the information regarding the replication package in Section~\ref{sec:conclusion}.

\section{Background}
\label{sec:background}

\subsection{Code Models}
\label{subsec:code_models}

The success of large language models in natural language processing, such as BERT~\cite{bert}, RoBERTa~\cite{RoBERTa}, and T5~\cite{T5}, has spurred the development of code-specific models like CodeBERT~\cite{CodeBERT}, GraphCodeBERT~\cite{GraphCodeBERT}, and CodeT5~\cite{wang2021codet5}. 
Trained on extensive, publicly available source code datasets~\cite{husain2019codesearchnet,javacorpus,py150}, these models have achieved state-of-the-art results in various software engineering tasks, including code completion~\cite{9794048}, program repair~\cite{10.1145/3510003.3510222}, and defect prediction~\cite{9462962}.

In general, the input to the code model consists of a sequence of tokens,
denoted by $x_1, \cdots, x_i$.
The model generates a probability distribution, $f_\theta (y_1 | x_1, \cdots, x_i)$,
which represents the likelihood of the next token in the sequence being $y_{1}$.
In particular, CodeGPT  uses \textit{beam-search}~\cite{beam-search} to
generate the next token.
The beam-search algorithm selects the top $k$ most probable tokens as the
initial beams, where $k$ is the beam size.
Then, the algorithm expands each of the $k$ beams by considering all possible
next tokens $y_{2}$, given the partial sequence $y_1, y_2$ and the input
sequence $x_1, \cdots, x_i$. The model calculates the joint probability of the
new sequences as:
\begin{equation*}
p(y_1, y_{2} | x_1, \cdots, x_i) = p(y_1 | x_1, \cdots, x_i) \times p(y_{2} | x_1, \cdots, x_i, y_1)
\end{equation*}
This iterative process perseveres until the attainment of a pre-established maximum target sequence length.
Upon fulfillment of the stopping criterion, the algorithm discerningly selects the beam with
the highest overall probability as the final generated target sequence.

It is important to note that our study does not directly engage with state-of-the-art models like OpenAI's Codex or ChatGPT due to the inaccessibility of their training data and potential legal issues related to testing these commercial systems. 
Instead, we focus on CodeGPT~\cite{CodeXGLUE}, a widely used code model with publicly available training datasets.

\subsection{Motivation}
\label{subsec:motivation}

Membership Inference Attack (MIA) poses a significant privacy risk by determining if a specific data point was used in training a Deep Neural Network (DNN) model~\cite{7958568}. Evaluating the vulnerability of code models to such attacks is essential for protecting the information of models and their training data. This section outlines the motivation for studying MIA on code models.

\vspace*{0.2cm}
\noindent \textbf{Motivation 1:}
MIA may bring threats of privacy leakage.
It is imperative to acknowledge privacy as a pivotal non-functional requirement from a developer's viewpoint in the developmental process of code models.
Code models are trained using a variety of data sources, ranging from publicly available data, such as open-source projects on GitHub, to more private and confidential data from companies.
This training data can encompass sensitive elements such as passwords, critical software implementation logic, and API keys~\cite{basak2023secretbench}.
Research indicates that language models have the potential to memorize and inadvertently reveal parts of their training data, including sensitive information~\cite{carlini21extracting,zheng-jiang-2022-empirical,yang2023memorzation}.
A malicious actor, through the utilization of MIA, can potentially ascertain whether a code model has been trained on datasets containing sensitive or confidential information, thus enabling further exploitative attacks aimed at data theft or sensitive information extraction.

\vspace*{0.2cm}
\noindent \textbf{Motivation 2:}
MIA may bring security threats. If a model's training data includes code with known or unknown security vulnerabilities~\cite{DBLP:conf/sp/PearceA0DK22}, the model may propagate such vulnerabilities to the systems that use the code generation models during their development. In this case, attackers may leverage MIA to identify what vulnerabilities are potentially included in the code models, and then launch further attacks on the systems that use the code models, putting these systems at risk.

\vspace*{0.2cm}
\noindent \textbf{Motivation 3:}
A further motivation for assessing MIA in code models lies in its capability to safeguard intellectual property.
Studies have indicated that open-source developers might not explicitly provide consent to data collectors for model training using their code~\cite{feitelson2021we}, a practice termed as \textit{unauthorized training}~\cite{CoProtector}.
Additionally, certain open-source codes possess licenses that preclude their utilization in model training.
MIA can serve as a tool to ascertain the inclusion of such protected codes in the training process.

\subsection{Threat Model}
\label{sec:threatmodel}

A threat model encompasses the assumptions regarding the positions and capabilities of both the attacker and the defender, along with a detailed description of the attack process.
In this study, we adopt the following assumptions to constitute our threat model.

\vspace*{0.2cm}
\noindent \textbf{Assumption 1 (Model Usage):}
We assume that the users of code models are afforded \textit{black-box} access, allowing them to interact with the models multiple times to accumulate pairs of inputs and outputs.

\vspace*{0.2cm}
\noindent \textbf{Assumption 2 (Model Parameters):}
We postulate that the attacker is restricted from accessing the model's parameters or gradient information, aligning with real-world scenarios where model owners typically offer their models as services accessible via APIs.
This service-oriented access restricts users to querying the model without direct exposure to underlying parameters or gradients.
For instance, OpenAI facilitates API access to its code completion services, aligning with this assumption.\footnote{\url{https://platform.openai.com/docs/introduction}}
Such an assumption is consistent with the premises adopted in previous works evaluating threats against code models~\cite{alert,MHM,you-see}.

\vspace*{0.2cm}
\noindent \textbf{Assumption 3 (Training Data):}
Following the baselines~\cite{hisamoto-etal-2020-membership}, we also assume that users may have access to portions of the models' training data.
Prevalent powerful code models are predominantly trained utilizing extensive open-source datasets.
For instance, Copilot~\cite{codex} undergoes training with natural language text and source code extracted from publicly accessible sources, such as code housed in public repositories on GitHub.
Similarly, CodeWhisperer enhances its performance through training on a substantial volume of publicly available code~\cite{AWSCodeWhisperer}.
The training data employed by well-known open-source code models, for example, CodeSearchNet~\cite{husain2019codesearchnet}, is also publicly accessible.
However, model owners might also employ their \textit{private} data for training purposes, keeping it inaccessible to users. This makes it plausible to assume that users can access only certain segments of the models' training data.

\section{Methodology}
\label{sec:methodology}

\begin{figure*}[!t]
  \centering
  \includegraphics[width=0.9\linewidth]{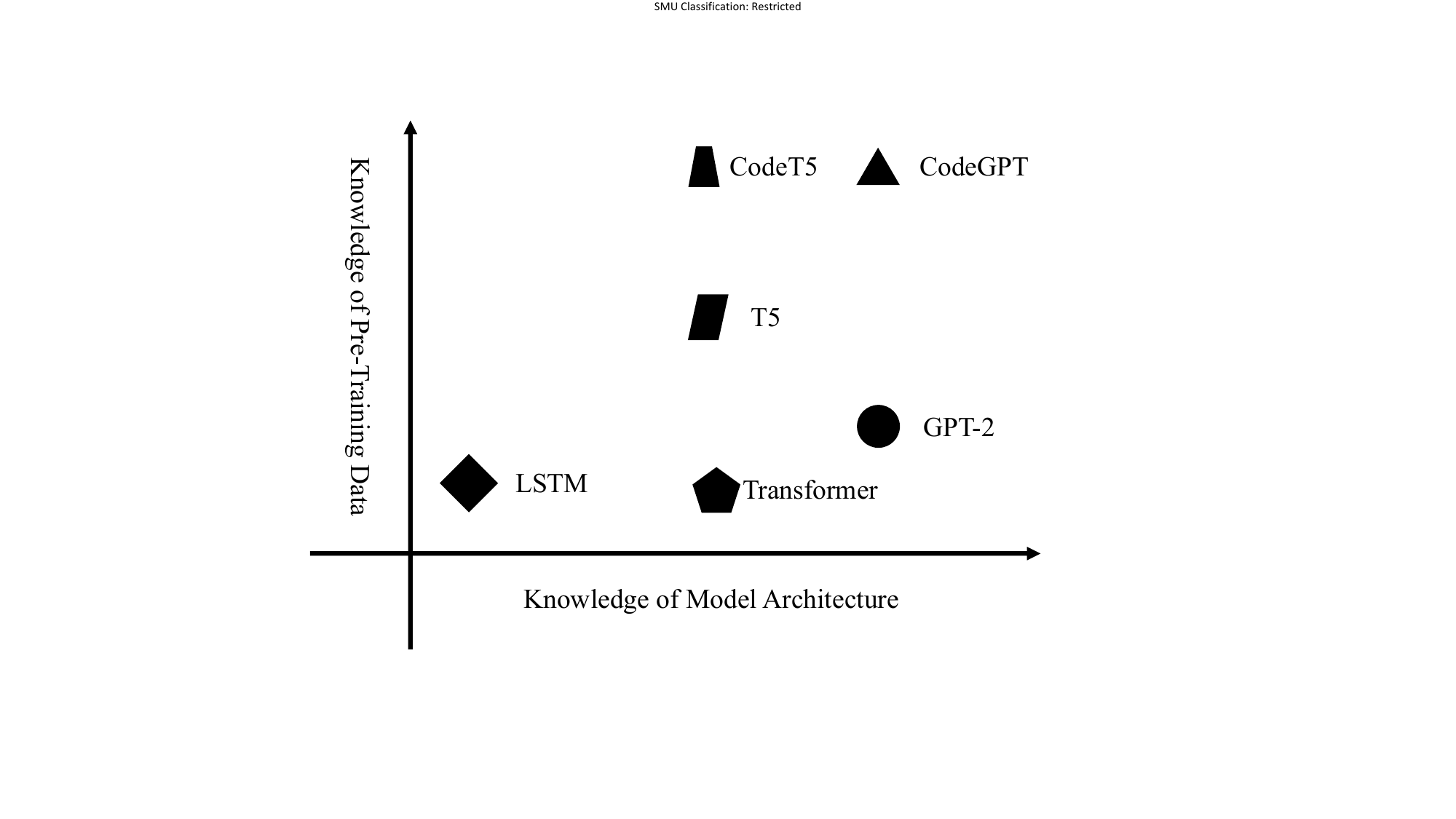}
  \caption{The overview of the proposed method. To achieve a high attack performance, our method coverts three types of information (including the ground truth, model input, and model output) into code embeddings and build a classifier on top of the embeddings.}
  \label{fig:overview}
\end{figure*}

This section explains our methodology. The overview of our method is shown in Figure~\ref{fig:overview}.
We first formulate the task and explain the design of our proposed approach \toolname.

\subsection{Task Formulation}

Existing membership inference attack (MIA) methods have exhibited satisfactory performance on classification models~\cite{7958568}.
However, applying these methods to code completion problems presents a markedly greater challenge.
Code completion can be conceptualized as a sequence of interconnected classifications, which substantially amplifies the complexity of addressing the MIA problem.

We formally define the MIA on code completion models as follows.
Let us consider a code completion model $\mathcal{M}$ and it is fine-tuned on a dataset $\mathcal{D}_{in}$.
The training dataset $\mathcal{D}_{in} = \{(x_i, y_i)\}_{i=1}^n$, where $x_i$ is the input, $y_i$
is the output, and $n$ is the size of the training set.
Following existing studies on MIA and other threats to code models~\cite{alert,you-see,9825895}, we assume that the model $\mathcal{M}$ is \textit{static}, i.e., the model does not change over time when the MIA is conducted.
The model $\mathcal{M}$ can be queried and complete code in a black-box manner:
$\hat{y}=\mathcal{M}(x)$; $\hat{y}$ is the output of the model $\mathcal{M}$
given the input $x$.
The attacker aims to build a binary classifier $\mathcal{G}$ to infer whether
an example $(x, y)$ is a member of the training set $\mathcal{D}_{in}$. The
goal of this classifier is:
\begin{equation*}
  \mathcal{G}(x,y,\hat{y})=
\left\{
    \begin{array}{ll}
        1 & \text{if~} (x,y) \in \mathcal{D}_{in}; \\
        0 & \text{otherwise}. \\
    \end{array}
\right.
\end{equation*}

\subsection{Training Surrogate Models}

Our proposed methodology, \toolname, operates in two steps.
Initially, we train a \textit{surrogate model} using part of the training data as the target victim model, but without direct interaction or access to the victim model's outputs.
Subsequently, in the second phase, the surrogate model undergoes queries utilizing its training data, as well as previously unseen non-training data.
The outcomes of these queries facilitate the training of a membership inference classifier, tasked with deducing the membership within the surrogate model's training set.

\begin{figure}[!t]
  \centering
  \includegraphics[width=0.8\linewidth]{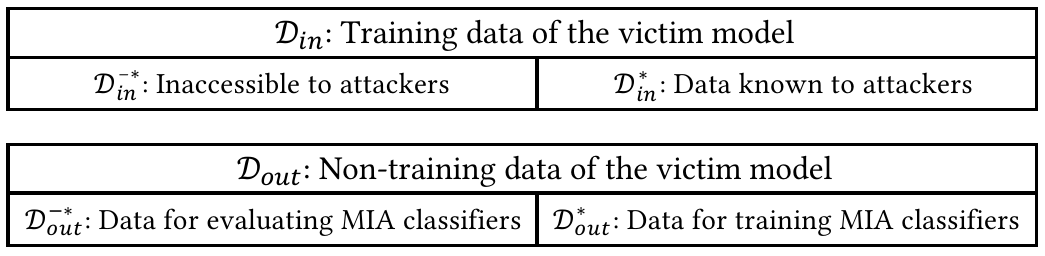}
  \caption{Splitting the datasets to train and evaluate surrogate models as well as the membership inference classifiers.}
  \label{fig:data-split}
\end{figure}

Figure~\ref{fig:data-split} shows how we split the dataset to train and
evaluate the proposed method.
Let $\mathcal{D}_{in}$ be the training set of the victim model.
As explained in the threat model in Section~\ref{sec:threatmodel}, we make a
practical assumption that the attacker can access a part of the training data,
which is denoted by $\mathcal{D}_{in}^{*}$.
The remaining training data is inaccessible to the attacker, which is denoted by $\mathcal{D}_{in}^{-*}$.
The attacker uses $\mathcal{D}_{in}^*$ to train a surrogate model $\mathcal{S}$.

$\mathcal{D}_{in}^{*}$ is also then used as the positive examples (i.e., as they are used to train the surrogate model) to train the membership classifiers.
But training a binary classifier requires both positive and negative examples.
So the attacker then finds another dataset $\mathcal{D}_{out}^*$ that are not used to train the surrogate model and the victim model.
It is a common practice to balance the training data (i.e., the ratio of positive and negative examples is $1:1$) to train a binary classifier, so we set $\mathcal{D}_{in}^{*}$ and $\mathcal{D}_{out}^*$ to be the same size. 
For each example $(x, y)$ in $\mathcal{D}_{in}^* \cup \mathcal{D}_{out}^*$, the attacker queries the surrogate model and obtains the corresponding output $\hat{y}_\mathcal{S} = \mathcal{S}(x)$.
The attacker creates a new dataset $\mathcal{D}_{MIA}$.
The input to the membership classifier is a tuple $ \langle x, y, \hat{y}_\mathcal{S} \rangle $, and the output is the membership label.
Formally, the label is decided as follows:
\begin{equation*}
  label =
  \left\{
    \begin{array}{ll}
        1 & \text{if~} (x,y) \in \mathcal{D}_{in}^*; \\
        0 & \text{if~} (x,y) \in \mathcal{D}_{out}^*. \\
    \end{array}
\right.
\end{equation*}
Note that as shown in Figure~\ref{fig:data-split}, the attacker can create another unseen dataset $\mathcal{D}_{out}^{-*}$ that shares the same size as $\mathcal{D}_{in}^{-*}$, to evaluate the MIA classifiers on the victim model.

In this paper, we consider multiple surrogate models to scrutinize the impact of the attacker's knowledge of the victim model on the efficacy of the attack.
The breadth of the attacker's knowledge concerning the victim model can be delineated across two dimensions: architectural comprehension and awareness of the pre-training data.
We incorporate four distinctive surrogate models within our analysis: (1) LSTM, (2) Transformer, (3) GPT-2, and (4) CodeGPT.

Employing CodeGPT as the surrogate model epitomizes the most favorable scenario for the attacker, who is presumed to know both the architecture and the pre-training data of the victim model.
Utilization of GPT-2 signifies a scenario where the attacker knows the victim model's architecture but lacks knowledge of the pre-training data.
Implementing the Transformer model signifies that the attacker possesses partial awareness of the victim model's architectural design\footnote{CodeGPT is a Transformer-based model.} but remains uninformed about the pre-training data.
The employment of LSTM represents a condition where the attacker lacks knowledge pertaining to both the architecture and the pre-training data of the victim model.

\subsection{Training MIA Classifiers}

We train an MIA classifier, utilizing a dataset forged through the employment of the surrogate model.
Hisamoto et al.\cite{hisamoto-etal-2020-membership}, while conducting MIA on natural language models, extract statistical features, constructing a classifier upon these foundational elements, incorporating aspects such as 1- to 4-gram precision and smoothed sentence-level BLEU score.
However, findings from prior research~\cite{hisamoto-etal-2020-membership} indicate limited effectiveness of these methods when applied to MIA on completion models.
In our exploration, we draw inspiration from the realization that pre-trained code models (e.g., CodeBERT~\cite{CodeBERT}) exhibit a commendable performance in classifying code inputs across varied categories, such as vulnerability assessments and code clone detection.
Motivated by this, we employ CodeBERT for the extraction of code embeddings, proceeding thereafter to train a classifier using these refined embeddings.

\subsubsection{Architecture Design and Input Processing}
By querying the surrogate model, we obtain its output that corresponds to the input.
A series of empirical studies~\cite{9609166,10.1145/3533767.3534390} have shown the effectiveness of pre-trained language models in code classification tasks, e.g., clone detection and defect prediction.
In this study, we use CodeBERT~\cite{CodeBERT}, which is one of the most widely used pre-trained code models.
To be more specific, we use three CodeBERT models to extract three code embeddings from the input, model output, and ground truth, respectively.
CodeBERT treats its input as a sequence of tokens.
Assuming that the sequence length is $L$, CodeBERT produces an embedding matrix of size $L \times 768$.

Typically, a classification model expects its input size to be fixed or it performs some processing to make the input of the same size.
In this study, we use the \textit{average pooling} method to compress the embedding metrics into a single embedding of size $768$.
The average pooling method is also used in previous studies~\cite{PTM4TAG,he2023representation}.
We conduct the same processing on the embeddings for the model output and the ground truth to obtain two $768$-dimensional vectors.
The processed embeddings are then concatenated together to form the final input.
We use a two-layer fully connected network with 768 hidden units and the \texttt{tanh} activation function as the MIA classifier.

\subsubsection{Model Training and Inference}
The proposed approach, \toolname, consists of the CodeBERT models and the MIA classifier; we denote their parameters with $\theta_p$ and $\theta_{c}$, respectively.
Given an example $\langle x, y, \hat{y_{\mathcal{S}}} \rangle$, we denote its label for the classification task as $l \in \{0,1\}$, where $1$ means the example is in the training set of the victim model and 0 otherwise.
$\mathcal{G}(\langle x_i, y_i, \hat{y_{\mathcal{S}}} \rangle)$ represents the prediction results of the MIA classifier.
We train the MIA classifier by optimizing the following loss function:
\begin{equation*}
  \arg \min_{\theta_p, \theta_f} \sum_{x_i \in D^*} \log (\mathcal{G}(\langle x_i, y_i, \hat{y_i} \rangle))
\end{equation*}
In the above equation, we train the attacker using a balanced dataset $D^*=\mathcal{D}_{in}^* \cup \mathcal{D}_{out}^*$, where $\mathcal{D}_{in}^*$ contains examples in the training data of the victim model and $\mathcal{D}_{out}^*$ contains unseen examples of the victim model.
To optimize the above loss function, we use back-propagation to update both parameters $\theta_p$ and $\theta_c$.

When training the MIA classifier, we use the output from the surrogate model.
However, in the inference stage (i.e., when an attack is performed), we use the output from the victim model to evaluate the attack performance.
We denote the output from the victim model as $\hat{y_v}$.
We send $\langle x, y, \hat{y_v} \rangle$ to the MIA classifier to obtain the prediction results $\mathcal{G}(\langle x, y, \hat{y_v} \rangle)$.
Note that training a DNN-based MIA classifier has a randomness that may affect the attack performance.
To mitigate the threats due to the randomness, we train the MIA classifier for three times using different random seeds to initialize the model parameters.
In our experiment, we report the average performance of the three trained MIA classifiers.

\subsection{Evaluation Metrics}

Aligned with preceding research~\cite{hisamoto-etal-2020-membership}, we assess the performance of MIA classifiers employing specific metrics: the attacker's power, quantified by the True Positive Rate (TPR), and the attacker's error, measured by the False Positive Rate (FPR).
Supplementing these metrics, we also incorporate the Area Under the Receiver Operating Characteristic (ROC) Curve (AUC-ROC).

\subsubsection{True Positive Rate (TPR)}
The True Positive Rate (TPR) represents the attacker's proficiency in accurately identifying instances that genuinely belong to the training dataset.
True positive rate can be considered as `attacker's power.'
A heightened TPR means that the attacker is strong in pinpointing these instances, thereby posing a potential risk to the confidentiality of the training data.

\subsubsection{False Positive Rate (FPR)}
The False Positive Rate (FPR) delineates the frequency at which the attacker erroneously categorizes instances as belonging to the training dataset when they do not.
False positive rate can be considered as `attacker's error.'
An elevated FPR suggests a lack of precision in the attacker's identifications, resulting in numerous false alarms and a diminished threat to the privacy of the training data.

\subsubsection{Area Under the ROC Curve (AUC)}
The Area Under the ROC Curve (AUC) manifests as a singular numeric value extracted from the ROC curve, encapsulating the overall performance of the attacker.
An AUC value proximate to 1 unveils a highly competent attacker, whereas a value nearing 0.5 implies that the attacker's performance is akin to random guess.
Superior AUC values signify enhanced performance by the attacker, while inferior values indicate subpar execution.

\section{Experiment Settings}
\label{sec:exp}

This section describes the configurations of our experiments, encompassing aspects such as the victim model, datasets, baseline methodologies, and implementation particulars.

\subsection{Victim Model}
\label{subsec:model}
Our study does not directly involve state-of-the-art models like OpenAI's Codex or ChatGPT.
We do not pick them as what data is used to train these models is unknown, making it infeasible to correctly evaluate the membership leakage risk.
Moreover, our goal is to evaluate the membership leakage risk, not to create an attack that can be operationalized by real adversaries to commercial services, which may raise potential legal issues.

As a result, we choose CodeGPT, a popular open-source code completion model included in the CodeXGLUE benchmark, as our main experiment subject to gain a deeper understanding of data privacy issues of code models.
CodeGPT consists of 12 layers of Transformer decoders, sharing the same model architecture and training objective of GPT-2~\cite{gpt-2}.
Different versions of CodeGPT models are released on the HuggingFace platform.\footnote{\url{https://huggingface.co/microsoft/CodeGPT-small-java}}
We use the `\texttt{microsoft/CodeGPT-small-java}' model, which is pre-trained on the Java code (around 1.6 million Java methods) in the CodeSearchNet dataset~\cite{husain2019codesearchnet}.
This model is pre-trained with randomly initialized model parameters.
Then, following the practice adopted in CodeXGLUE benchmark~\cite{CodeXGLUE}, we further fine-tune CodeGPT-small-java on a subset of 1\% randomly sampled examples from JavaCorpus to obtain the victim model. Note that JavaCorpus and CodeSearchNet are two different datasets and JavaCorpus is not included in the “pre-training dataset” of CodeGPT-small-java.

Additionally, to further evaluate the generalizability of our proposed membership inference attack, we consider five open-source models: \texttt{CodeGen}~\cite{codegen}, \texttt{CodeParrot}~\cite{codeparrot}, \texttt{gpt-neo}~\cite{gpt-neo}, \texttt{PolyCoder-160M}, and \texttt{PolyCoder-0.4B}~\cite{10.1145/3520312.3534862}.
These models are widely used as experiment subjects in recent studies~\cite{stupidbug,yang2023memorzation,10.1145/3520312.3534862}.
\texttt{CodeGen} models~\cite{codegen} adopt a standard transformer decoder with left-to-right causal masking.
We choose the \texttt{CodeGen-multi-350M} model, which is pre-trained on the BigQuery dataset and can support Java code completion.
\texttt{CodeParrot}~\cite{codeparrot} adopts the GPT-2 architecture with 1.5 billion parameters.
We choose two variants of PolyCoder models: \texttt{PolyCoder-160M} and \texttt{PolyCoder-0.4B}, with 160M and 0.4B parameters, respectively.
For \texttt{gpt-neo}~\cite{gpt-neo}, we use its 125M parameter version.

\subsection{Datasets}
In this study, we use datasets for code completion as it is one of the
most important tasks in software engineering.
Given a piece of code snippet, the goal of code completion is to predict the next tokens or lines.
We consider a popular dataset included in the CodeXGLUE benchmark~\cite{CodeXGLUE}: JavaCorpus~\cite{javacorpus}.
Allamanis and Sutton collect the JavaCorpus dataset~\cite{javacorpus}, containing over 14,000 Java projects from GitHub.
The CodeXGLUE benchmark follows the settings in Karampatsis et al.'s study~\cite{big_code} and samples 1\% of the subset from the JavaCorpus dataset, ending up with 12,934/7,189/8,268 files for the training/validation/test set, respectively.
Then, the CodeXGLUE benchmark preprocesses the dataset by tokenizing the source code using a Java parser and removes all the comments.
As reported in the paper~\cite{CodeXGLUE}, strings that are longer than 15 characters are replaced with empty strings.

We further explain how we split the dataset to train and evaluate the MIA classifiers.
For victim models, their training and testing dataset have 12,934 and 8,268 examples, respectively. We randomly sample some examples (for example 10\%, i.e., 1,293) from the training set; the sampled examples will be used to train the surrogate model and then used as the positive examples (i.e., ground truth label is 1) to train MIA classifiers. As training MIA classifiers also requires negative examples whose ground truth labels are 0, we randomly sample the same number of examples from victim models' testing set as the negative examples to train MIA classifiers. Similarly, we need to construct positive and negative examples to evaluate MIA classifiers. We randomly sample 1,293 examples (10\% of the whole training set) from both the remaining training set and the remaining testing set. This strategy ensures that there is no overlap between the training and evaluation data of MIA classifiers.

\begin{table}[!t]
  \centering
  \caption{Statistics of dataset for training and evaluation models. The training set of surrogate models is randomly sampled from the training set of the victim model, which is also used as the positive examples to train MIA classifiers.}
  \begin{tabular}{r|l|l}
    \toprule
    \multirow{2}{*}{Model} & \multicolumn{2}{c}{Data size} \\ \cmidrule(lr){2-3}
    & Training & Testing \\ \midrule
    Victim Model & 12,934 & 8,268 \\ \hline
    Surrogate Model & 1,293 & 8,268 \\ \hline
    MIA Classifiers & 2,586 & 2,586 \\ \hline
    \toprule
  \end{tabular}
  \label{tab:data_statistic}
\end{table}

\begin{table}[!t]
  \centering
  \caption{Model names used in the study and their corresponding names on the HuggingFace platform.}
  \begin{tabular}{r|l}
    \toprule
    Model & Model Name on HuggingFace \\ \midrule
    \texttt{CodeGPT}    & \texttt{microsoft/CodeGPT-small-java}   \\ \hline
    \texttt{CodeGen}    & \texttt{Salesforce/codegen-350M-multi}   \\ \hline
    \texttt{CodeParrot}    & \texttt{codeparrot/codeparrot-small}  \\ \hline
    \texttt{PolyCoder-160M}    & \texttt{NinedayWang/PolyCoder-160M}   \\ \hline
    \texttt{PolyCoder-0.4B}    & \texttt{NinedayWang/PolyCoder-0.4B}  \\ \hline
    \texttt{gpt-neo}    & \texttt{EleutherAI/gpt-neo-125m}  \\ 
    \toprule
  \end{tabular}

  \label{tab:namemap}
\end{table}

\subsection{Baselines}

This study utilizes the techniques used for evaluating natural language models.
Although MIA is an important threat to AI and has been studied since 2017, researchers mainly focus on classification tasks and there are only a few studies on attacking generative language models.
While these baselines are not specifically designed for code models, there is a lack of membership inference attacks tailored for code completion tasks.
Therefore, we adapt existing approaches by Hisamoto et al.~\cite{hisamoto-etal-2020-membership} and Carlini et al.~\cite{carlini21extracting}, originally designed for natural language models, to serve as baselines for our investigation into membership information leakage risk in code models.
These two lines of methods are categorized as feature-based Classification and metrics-based ranking methods. Our proposed method is different from their methods. Instead of computing manually defined features from examples, we infer membership by considering the input example, ground truth, and how models react (i.e., completion) on the example. Specifically, we design a novel membership inference classifier that considers the three types of information together and show that each information matters for the final prediction.

\subsubsection{Feature-based Classification}
Our first baseline is a set of classifiers trained by features.
Hisamoto et al.~\cite{hisamoto-etal-2020-membership} develop MIA on a machine translation system.
They extract some features from the model output and the ground truth to build a binary classifier.
The considered features are modified 1- to 4-gram precision and smoothed sentence-level BLEU score~\cite{10.3115/1218955.1219032}.
The intuition is that if an unusually large number of n-grams in $y$ matches $\hat{y}$, then it could be a sign that this is in the training data and the victim model memorizes it.
Hisamoto et al. try different types of classifiers.
Following their settings, our study uses Nearest Neighbors (NN), Decision Tree (DT), Naive Bayes (NB), and Multi-layer Perceptron (MLP).
Additionally, we consider deep neural networks (DNN).

\subsubsection{Metrics-based Ranking}
\label{subsubsec:metrics}
Second, we utilize ranking techniques based on certain metrics as another baseline.
Carlini et al.~\cite{carlini21extracting} investigate the data extraction attack on language models.
Specifically, they propose a process that involves sampling numerous examples from a language model and subsequently ranking them using specific metrics.
The objective is to rank examples from the training dataset in the top positions, which closely aligns with the goal of MIA. We refer to this research approach as \textit{metrics-based ranking} for MIA.
After ranking the examples using different metrics, the attacker needs to set a \textit{cut-off} position to determine what examples will be considered as in the training set.
In Carlini et al.'s work~\cite{carlini21extracting}, the cut-off position is the top 10\% of the examples.
Our study uses a balanced dataset to evaluate MIA, so we set the cut-off position at 50\%.
By doing so, we can assign the predicted labels to each example and compute the evaluation metrics like power, error, and AUC.
Some metrics in Carlini et al.'s work is designed for natural language, e.g., converting all the characters to lowercase, which is not suitable for code models.
So this paper considers the following metrics to infer data membership in code models.

\vspace{0.2cm}
\noindent \textbf{Perplexity.}
Perplexity~\cite{ppl} is a measurement of how well a probability model predicts a sample.
The logarithm of perplexity is the formally defined as $log(P) = -\frac{1}{N} \sum_{i=1}^{N} \log P(w_i \mid w_1, w_2, \ldots, w_{i-1})$, where $N$ is the total number of words in a test example, $w_i$ is the $i$-th word, and $P(w_i \mid w_1, w_2, \ldots, w_{i-1})$ is the conditional probability of the $i$-th word given the previous words in the sequence.
A low perplexity signifies that the model is good at predicting the sample.
Intuitively, a low perplexity for a specific example can suggest that the model has previously encountered this example during training.
We use the victim model to compute the perplexity of each example and rank them in ascending order.

\vspace{0.2cm}
\noindent \textbf{Comparing perplexity of another language model.}
As described by Carlini et al.~\cite{carlini21extracting}, this metric is computed by the ratio of log-perplexities of the victim model and another language model.
In this paper, we use the surrogate model as the second model.
So the metric is computed as $\frac{log(P_v)}{log(P_s)}$, where $P_v$ is the perplexity computed using the victim model and $P_s$ is the perplexity computed using the surrogate model.
We rank examples in descending order.

\vspace{0.2cm}
\noindent \textbf{Comparing to \texttt{zlib} compression.}
Carlini et al.~\cite{carlini21extracting} also consider the \texttt{zlib} compression.
When using \texttt{zlib}~\cite{zlibnet} to compress a sequence of tokens, \texttt{zlib} represents the compressed sequence using as few bits as possible.
The \texttt{zlib} entropy of a sequence is defined as the number of bits used to represent the compressed sequence.
The attacker uses $\frac{log(P_v)}{zlib}$, i.e., the ratio of the victim model perplexity and the \texttt{zlib} entropy as a membership inference metric.

\subsection{Implementation and Experiment Platforms}
\label{subsec:implementation}
We utilize the replication package provided in the CodeXGLUE benchmark~\cite{CodeXGLUE} to fine-tune CodeGPT.
However, as Hisamoto et al.'s paper~\cite{hisamoto-etal-2020-membership} does not provide a replication package, we follow the methodology and guidelines described in their paper to implement the MIA classifiers based on the statistical features.
To implement the decision tree, we use GINI impurity for the splitting metrics and the max depth is set as 5.
Naive Bayes uses Gaussian distribution.
We set the number of neighbors to 5 and use Minkowski distance to implement the NN classifier.
For MLP, we set the size of the hidden layer to be 50, the activation function to be ReLU, and the $L_2$ regularization term $\alpha$ to be 0.0001.
The hyperparameters settings follow the settings of Hisamoto et al.'s study~\cite{hisamoto-etal-2020-membership}.
To evaluate the effectiveness of metrics-based ranking methods, we leverage the replication package provided by Carlini et al.~\cite{carlini21extracting}.\footnote{\url{https://github.com/ftramer/LM_Memorization}}

We perform our experiments on a computer running Ubuntu 18.04 with 4 NVIDIA GeForce A5000 GPUs.
To mitigate the effect of randomness in training MIA classifiers, we repeat each experiment using three different random seeds for model parameter initialization in each run.
We compute the average results for each evaluation metric, which enable us to provide a more accurate and reliable representation of the MIA classifier's performance, which is less susceptible to the influence of random factors.

\section{Results}
\label{sec:result}

In this section, we evaluate the risk of membership leakage in code models by answering the following three research questions (RQs):
\begin{itemize}[leftmargin=*]
  \item \textbf{RQ1.} \textit{To what extent are code models vulnerable to membership inference attacks?}
  \item \textbf{RQ2.} \textit{What are the factors affecting the membership leakage risk?}
  \item \textbf{RQ3.} \textit{What are the features of the training examples whose memberships are more likely to be correctly inferred?}
\end{itemize}

In the first RQ, we apply our proposed approach and two baselines~\cite{carlini21extracting,hisamoto-etal-2020-membership} to CodeGPT to evaluate the risks exposed by these attacks.
Then, in the second RQ, we conduct the attack in different settings to simulate the different prior knowledge the attackers (e.g., model architecture, size of known training data, etc.) have and analyze the factors that affect the membership leakage risk in code models.
Lastly, we investigate the features of the training examples whose membership is more likely to be correctly inferred.

\subsection*{RQ1. To what extent are code models vulnerable to membership inference attacks?}

In this question, we evaluate our proposed approach \toolname and baseline attacks~\cite{carlini21extracting,hisamoto-etal-2020-membership} on CodeGPT model~\cite{CodeXGLUE}.
As mentioned in the threat model, it is reasonable to assume that the attacker can access part of the training data of the victim model.
In this RQ, we assume that the attacker knows $20\%$ of the training data of the victim model, which is used to train the surrogate model and the MIA classifier to infer the data membership.
In this experiment, the surrogate model is a pre-trained CodeGPT model and then fine-tuned on part of the victim model's training data that are known to the attacker.
In the following RQ, we will try different surrogate models to analyze the impact of the surrogate model architecture on the risk of privacy attacks.

\begin{table}[!t]
  \centering
  \caption{The performance of different membership inference attacks on the CodeGPT model. \textit{w.o.} means `without', i.e., we exclude a certain part from the proposed method. H-Attack and C-Attack refer to the works by Hisamoto et al.~\cite{hisamoto-etal-2020-membership} and Carlini et al.~\cite{carlini21extracting}.}
  \begin{tabular}{l@{\hskip 0.3in}lS[table-format=2.2]S[table-format=2.2]S[table-format=1.2]}
    \toprule
    \textbf{} & \textbf{Variants} & \multicolumn{1}{c}{\textbf{Power}} & \multicolumn{1}{c}{\textbf{Error}} & \multicolumn{1}{c}{\textbf{AUC}} \\ \midrule
    \multirow{4}{*}{\rotatebox[origin=c]{90}{Ours}} & \toolname         &  \textbf{ 0.95} & \textbf{ 0.10}  & \textbf{0.98} \\
                                     & \toolname \textit{w.o.} input  & 0.70 & 0.50  & 0.63\\
                                     & \toolname \textit{w.o.} truth  & 0.87 & 0.38  & 0.83\\
                                     & \toolname \textit{w.o.} output  & 0.65 & 0.27  & 0.60\\
    \addlinespace
    \multirow{5}{*}{\rotatebox[origin=c]{90}{H-Attack}} & Naive Bayes  & 0.23 & 0.17  & 0.58\\
                                     & Decision Tree     & 0.30 & 0.25 & 0.57 \\
                                     & Nearest Neighbor & 0.23 & 0.25 & 0.49 \\
                                     & Multi-layer Perceptron          & 0.28 & 0.22 & 0.58 \\
                                     & Deep Neural Network          & 0.21 & 0.27 & 0.58 \\
    \addlinespace
    \multirow{3}{*}{\rotatebox[origin=c]{90}{C-Attack}} & Perplexity          & 0.58 & 0.42  &  0.58 \\
                                     & Compare Perplexity &  0.47  & 0.53  & 0.47       \\
                                     & Compare \texttt{zlib}           & 0.55 & 0.45 & 0.55 \\ \bottomrule
    \end{tabular}
    \label{tab:attack-perf}
\end{table}

We consider two branches of baseline MIA methods: classification-based and metrics-based attacks.
Our proposed method \toolname and the work by Carlini et al.~\cite{carlini21extracting} are classification-based, which use a classifier to infer the data membership.
For classification-based attacks, we compute the performance metrics of the corresponding classifiers, including the accuracy, precision, recall, F1 score, and AUC.
Carlini et al.~\cite{carlini21extracting} try different metrics to infer the data membership.
More specifically, the data that are more likely to be the training data will be ranked in a higher position.

The experiment results are listed in Table~\ref{tab:attack-perf}, which presents the attacker's power (true positive rate), attacker's error (false positive rate), and AUC scores
for different attacks of different types of attacks used in the experiment.
The table shows that \toolname attack has the highest power score of 0.95, indicating that it is the most effective in identifying members in the training dataset.
The error score for \toolname is 0.10, indicating that it incorrectly identified some non-members as members.
The AUC score for \toolname is 0.98, indicating that the proposed approach is very effective.

The proposed method takes code embeddings of three parts: the input to the victim model, the ground truth, and the output from the victim model.
We conduct an ablation study to analyze the benefits of each part.
Table~\ref{tab:attack-perf} shows the results when we remove one part from \toolname's input.
For example, the row `\textit{w.o.} truth' shows the results when we remove the ground truth from the input.
The results show that all the three parts contribute to the effectiveness of the proposed method.
Excluding the input, ground truth, and model output will reduce the AUC scores by 0.35, 0.15, and 0.38, respectively.
It suggests that the model output is the most important part contributing to the effectiveness of the proposed method, followed by the input and the ground truth.

We also analyze the effectiveness of five classification-based attacks used by Hisamoto et al.~\cite{hisamoto-etal-2020-membership}: Decision Tree, Naive Bayes, Nearest Neighbor, Multi-Layer Perceptron, and Deep Neural Network.
Among them, the Decision Tree attack has the highest power score of 30.40, while Naive Bayes had the lowest power score of 22.98.
However, all three classification-based attacks had relatively low AUC scores below 0.6, indicating that they are less effective than the \toolname attack at identifying membership.
The AUC score of Nearest Neighbor is even lower, only 0.49.
In Table~\ref{tab:attack-perf}, the results for metric-based methods are obtained by setting the cut-off position as 50\%.
Under this setting, ranking using the three metrics (i.e., perplexity, comparing perplexity, and comparing \texttt{zlib}) achieves AUC scores of 0.58, 0.47, and 0.55, respectively.
The results show that both two baselines are less effective than our proposed approach at identifying members in the training dataset.

We further evaluate the proposed method on five additional large language models of code: \texttt{CodeGen}~\cite{codegen}, \texttt{CodeParrot}~\cite{codeparrot}, \texttt{gpt-neo}~\cite{gpt-neo}, \texttt{PolyCoder-160M}, and \texttt{PolyCoder-0.4B}~\cite{10.1145/3520312.3534862}.
We train each victim model for 5 epochs and apply the proposed method to each model.
We use the best configurations of our method \toolname from RQ1, i.e., using the CodeGPT model as the surrogate model and the attacker knowing 20\% of the training data.
We also apply the baseline to these models.
As the AUC score reflects the overall performance of a classifier, we report the AUC scores of the proposed method.
We run membership inference attacks on each model using 5 different random seeds.
We observe that the superior performance of \toolname can generalize to the five newly evaluated models.
\toolname achieves AUC scores of 0.92, 0.95, 0.93, 0.94, and 0.94 on \texttt{CodeGen}, \texttt{CodeParrot}, \texttt{gpt-neo}, \texttt{PolyCoder-160M}, and \texttt{PolyCoder-0.4B}, respectively.
In contrast, the AUC scores of both the feature-based classification and metric-based ranking baselines are lower than 0.60.
The results show that \toolname achieves better performance than baselines on the five additionally evaluated models.

\begin{tcolorbox}[boxrule=0pt,frame hidden,sharp corners,enhanced,borderline north={1pt}{0pt}{black},borderline south={1pt}{0pt}{black},boxsep=2pt,left=2pt,right=2pt,top=2.5pt,bottom=2pt]
  \textbf{Answers to RQ1}:    The evaluated victim model shows a \textbf{high risk} of leaking the training data membership information. The \toolname attack is the most effective at identifying members, with an AUC score of 0.98, highlighting the need for better safeguards to mitigate this risk.
\end{tcolorbox}

\subsection*{RQ2. What are the factors affecting the membership leakage risk?}
The previous RQ demonstrates that code models are vulnerable to membership inference attacks. 
Here, we examine the factors influencing membership leakage risk. 
Since baseline methods are not effective in inferring data membership, we only evaluate our proposed method, \toolname. 
We focus on analyzing four key factors.

\begin{enumerate}[leftmargin=*]
  \item \textbf{The training epochs of the victim model}. Previous research~\cite{8429311} on classification models suggests that the risk of membership information leakage increases with more training epochs. We train each victim model for 5 epochs and apply \toolname to 5 variants of the model and observe the attacker's power.
  \item \textbf{The surrogate models}. As explained in Section~\ref{subsec:model}, we use surrogate models of different architectures to simulate the attacker's prior knowledge of the victim model. Intuitively, a surrogate model that can better mimic the victim model will be more effective in the attack.
  We choose four surrogate models: CodeGPT, GPT-2, 12-Layer Transformer, and LSTM.
  \item \textbf{The victim models}. We evaluate the membership leakage risk of six open-source models: \texttt{CodeGPT}, \texttt{CodeGen}, \texttt{CodeParrot}, \texttt{gpt-neo}, \texttt{PolyCoder-160M}, and \texttt{PolyCoder-0.4B}.
  \item \textbf{The ratio of training data that is known to the attacker.}
  If the attacker knows more data that is used to train the victim model, the attacker may train a better surrogate model and MIA classifier, which may lead to a higher attack success rate.
  We evaluate using two settings: 10\% and 20\% of the training data are known to the attacker.
\end{enumerate}

To systematically evaluate the impact of these factors, we adopt the \textit{Design of Experiments (DoE)}~\cite{montgomery2009design} method.
The primary goal of DoE is to identify which factors (e.g., the choice of surrogate model in our study) most significantly affect the outcome of a process or system (e.g., the attacker's power).
DoE has been widely used in various engineering fields, including software engineering. 
For example, Cotroneo et al.~\cite{cotroneo2023vulnerabilities} apply DoE to understand the impact of different factors on data poisoning attacks for code models. 
Following their practice, we employ the \textit{full factorial design}, which considers all possible combinations of factors.
Specifically, we perform a total of 240 experiments (6 victim models $\times$ 5 epochs $\times$ 4 surrogate models $\times$ 2 ratios of known training data).
We conduct an \textit{Analysis of Variance (ANOVA)}~\cite{st1989analysis} to assess each factor's impact on the attacker's power. 
We determine a factor's importance by looking at the percentage of total variation it accounts for, known as the portion of Sum of Squares Total (\textbf{SST \%}). 
A factor is considered important if it explains a large part of the overall variation.
Each factor's significance also comes with a $p$-value. 
A $p$-value lower than 0.05 indicates that the factor has a significant impact on the attacker's power.

\begin{table}[!t]
  \centering
  \caption{ANOVA analysis of what factors have more impact on the attacker's power. We consider four factors: the surrogate models (\textit{Surro M}), the victim model (\textit{Victim M}), the ratio of known training data (\textit{Ratio}), and the training epochs of the victim model (\textit{Epoch}). The impact is measured by the proportion of total variation (SST \%) it can explain. A higher percentage indicates a more significant impact.}
  \begin{tabular}{lcc}
  \toprule
  \textbf{Factors} & \textbf{SST (\%)} & \textbf{$p$-value} \\
  \midrule
  \textit{Victim M} & 0.10\% & $>0.05$ \\
  \rowcolor{lightgray} \textbf{\textit{Surro M}} & \textbf{42.70\%} & \textbf{$<0.05$} \\
  \rowcolor{lightgray} \textbf{\textit{Ratio}} & \textbf{0.70\%} & \textbf{$<0.05$} \\
  \textit{Epoch} & 0.04\% & $>0.05$ \\
  \textit{Victim M} * \textit{Surro M} & 0.22\% & $>0.05$ \\
  \textit{Victim M} * \textit{Ratio} & 0.10\% & $>0.05$ \\
  \textit{Victim M} * \textit{Epoch} & 0.29\% & $>0.05$ \\
  \rowcolor{lightgray} \textbf{\textit{Surro M} * \textit{Ratio}} & \textbf{53.71\%} & \textbf{$<0.05$} \\
  \textit{Surro M} * \textit{Epoch} & 0.12\% & $>0.05$ \\
  \textit{Ratio} * \textit{Epoch} & 0.06\% & $>0.05$ \\
  \textit{Victim M} * \textit{Surro M} * \textit{Ratio} & 0.24\% & $>0.05$ \\
  \textit{Victim M} * \textit{Surro M} * \textit{Epoch} & 0.66\% & $>0.05$ \\
  \textit{Victim M} * \textit{Ratio} * \textit{Epoch} & 0.28\% & $>0.05$ \\
  \textit{Surro M} * \textit{Ratio} * \textit{Epoch} & 0.11\% & $>0.05$ \\
  \bottomrule
  \end{tabular}
  \label{tab:DoE}
\end{table}

Table~\ref{tab:DoE} shows the results of the ANOVA analysis.
The factors that have a significant impact on the attacker's power are highlighted in gray.
Notably, the surrogate model (\textit{Surro M}) alone accounts for 42.70\% of the total variation, highlighting its crucial role in influencing the attacker's power. Additionally, the interaction between the surrogate model and the ratio of known training data (\textit{Surro M} * \textit{Ratio}) explains an even larger portion of the variation at 53.71\%, suggesting that the combination of these two factors is particularly influential.
In contrast, other factors such as the victim model (\textit{Victim M}) and training epochs (\textit{Epoch}), as well as their interactions, show minimal impact with SST \% values under 1\% and $p$-values greater than 0.05, indicating they do not significantly affect the attacker's power.
Given the findings, we further conduct a series of statistical tests to validate the impact of each factor on the attacker's power in more detail.

\vspace{0.2cm}
\noindent \textbf{The training epochs of the victim model.}
We train CodeGPT for 10 epochs and apply \toolname to 10 variants of the model, each trained for 1, 2, 3, ..., 10 epochs.
We apply \toolname to each variant of the models and report the corresponding attacker's power.
For a victim model trained for $n$ epochs, we obtain 4 results as we use 4 different surrogate models (i.e., CodeGPT, GPT-2, Transformer, LSTM), denoted by $A_n$.
For each epoch (e.g., $i$ and $j$), we conduct a Wilcoxon signed-rank test~\cite{Wilcoxon} to compare the attacker's power $A_i$ and $A_j$ on the two models.
We make the following null hypothesis:

\begin{quote}
  \textit{There is no significant difference between the attacker's power on victim models trained for $i$ and $j$ epochs.}
\end{quote}

When multiple tests are performed, the probability of obtaining at least one false positive result increases~\cite{toothaker1993multiple}.
Following previous studies, we apply the Bonferroni correction~\cite{sedgwick2012multiple} to adjust the significance level for each individual test by dividing it by the number of tests being performed.
Our results show that the $p$-value of tests on each pair is larger than 0.05, failing to reject the null hypothesis.
Thus, the Wilcoxon test does not reveal any statistically significant differences between each pair of data sets.

We further compute the standard deviation of the MIA's precision and recall scores for the models trained for different epochs.
Taking precision scores as an example, for each model trained for $n$ epochs, we obtain $n$ values of the precision scores. We then compute the standard deviation of these $n$ values to measure the stability and consistency of the attack performance on models trained for various epochs.
A small standard deviation indicates that the attack performance is consistent.
A larger standard deviation indicates a greater difference in models' vulnerability to MIA under different training epochs settings.
The standard deviation values of the precision and recall are both less than $0.0005$, indicating that the number of training epochs of the victim model has little impact on the risk of privacy attacks.
Nonetheless, it also cautions model developers that even though these state-of-the-art models are trained for just a few epochs, they remain susceptible to such attacks.

\begin{figure}[!t]
  \centering
  \includegraphics[width=0.8\linewidth]{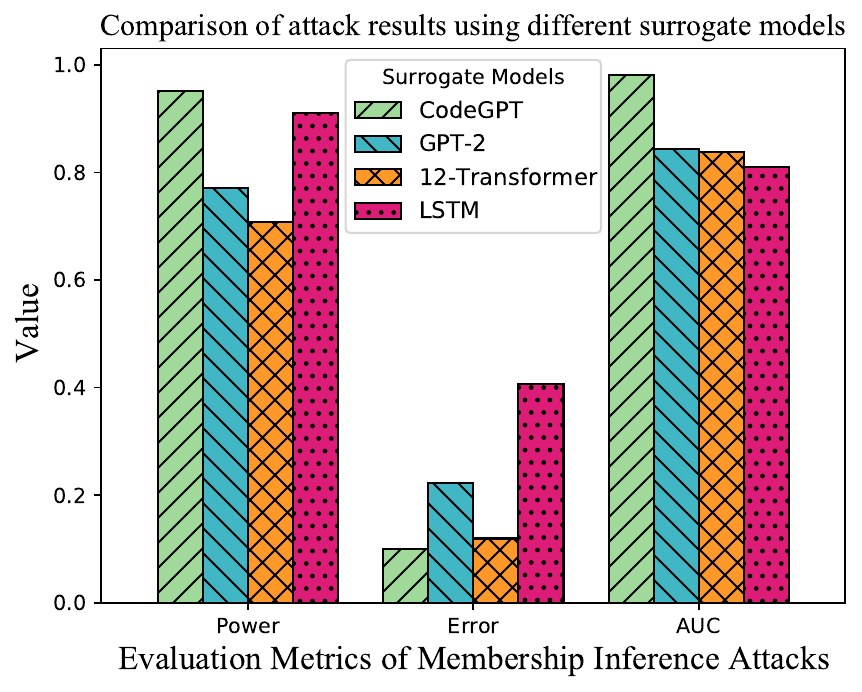}
  \caption{The impact of different choices of surrogate models on the attack performance.}
  \label{fig:vary-surrogate}
\end{figure}

\vspace{0.2cm}
\noindent \textbf{The surrogate models.}
We choose four surrogate models: CodeGPT, GPT-2, 12-Layer Transformer, and LSTM.
Figure~\ref{fig:vary-surrogate} shows the attack performance (i.e., attacker's power, error, and AUC) when using different surrogate models.
As can be seen in the table, using CodeGPT model can outperform the other surrogate models; it has the highest power of 0.95, the lowest error of 0.10, and the highest AUC of 0.98.
This indicates that the CodeGPT model, which shares the same architecture and pre-training data with the victim model, has the strongest attack performance among the surrogate models tested.

Notably, we observe a decreasing trend in the AUC scores of the GPT-2, 12-Layer Transformer, and LSTM models.
Although the LSTM model has a high power close to that of CodeGPT, its error rate is the highest (0.41) and the AUC is the lowest (0.81).
This implies that using the LSTM model as the surrogate is less effective than the other models in the attack.
In summary, the selection of a surrogate model greatly impacts the success of membership inference attacks.
More specifically, an attacker gains a considerable advantage when employing a surrogate model that closely resembles the victim model, e.g., knowing the architecture and pre-training data of the victim model.

\vspace{0.2cm}
\noindent \textbf{The ratio of training data that is known to the attacker.}
We evaluate using two settings: 10\% and 20\% of the training data are known to the attacker.
The results of our experiments are presented in Table~\ref{tab:vary-ratio}, which illustrates the attack performance under different ratios of known training data.
As demonstrated in the table, the attack performance varies across different models and ratios of known training data.

The specific impact of this increase varies depending on the model architecture and the performance metric considered.
For instance, while the CodeGPT and 12-Transformer models show a consistent improvement in attack success rates across all metrics as the proportion of known training data increases, the GPT-2 and LSTM models exhibit more nuanced behavior. The GPT-2 model experiences a decrease in attack success rates when more training data is known, while the LSTM model shows mixed results, with improvements in some metrics and declines in others.
Overall, 3 out of four models show an increase in attack success rates as the proportion of known training data increases.
In general, our results support the intuition that increasing the proportion of known training data leads to a higher attack success rate.

\begin{table}[!t]
  \centering
  \caption{Attack performances when the attacker knows different portion of the victim model's training data. The 10\% and 20\% mean the percentages of training data known to the attacker. The directions of the arrows indicate how the metrics change when the ratio of known training data increases: $\uparrow$ means the metric increases and $\downarrow$ means the metric decreases. The colors of the arrows indicate how the attack performance changes: blue means the attack performance improves (e.g., the attacker's power increases) and red means the attack performance degrades (e.g., the attacker's error increase).}
  \begin{tabular}{lcccccc}
  \toprule
  Model & \multicolumn{2}{c}{Power} & \multicolumn{2}{c}{Error} & \multicolumn{2}{c}{AUC} \\
  \cmidrule(lr){2-3} \cmidrule(lr){4-5} \cmidrule(lr){6-7}
        & 10\%      & 20\%      & 10\%      & 20\%      & 10\%    & 20\%    \\
  \midrule
  CodeGPT         & 0.87 & 0.95\textcolor{blue}{$\uparrow$} & 0.23 & 0.10\textcolor{blue}{$\downarrow$} & 0.89 & 0.98\textcolor{blue}{$\uparrow$} \\
  GPT-2            & 0.87 & 0.77\textcolor{red}{$\downarrow$} & 0.20 & 0.22\textcolor{red}{$\uparrow$} & 0.90 & 0.84\textcolor{red}{$\downarrow$} \\
  12-Transformer  & 0.63 & 0.71\textcolor{blue}{$\uparrow$} & 0.33 & 0.12\textcolor{blue}{$\downarrow$} & 0.70 & 0.84\textcolor{blue}{$\uparrow$} \\
  LSTM            & 0.74 & 0.91\textcolor{blue}{$\uparrow$} & 0.32 & 0.41\textcolor{red}{$\uparrow$} & 0.77 & 0.81\textcolor{blue}{$\uparrow$} \\
  \bottomrule
  \end{tabular}
  \label{tab:vary-ratio}
  \end{table}

\begin{tcolorbox}[boxrule=0pt,frame hidden,sharp corners,enhanced,borderline north={1pt}{0pt}{black},borderline south={1pt}{0pt}{black},boxsep=2pt,left=2pt,right=2pt,top=2.5pt,bottom=2pt]
  \textbf{Answers to RQ2}:   The number of training epochs of the victim model has little impact on the risk of membership leakage.
  However, the risk is higher if an attacker knows the victim model better, e.g., the model's architecture and training data.
\end{tcolorbox}

\subsection*{RQ3. What are the features of the training examples whose memberships are more likely to be correctly inferred?}
In this research question, we investigate the features of the training examples whose memberships are more likely to be correctly inferred.
We use the same experiment setup in RQ1.
We split the training examples into two groups: (1) successfully inferred examples and (2) unsuccessfully inferred examples.
We compute a list of features of these examples and compare the two groups of examples.
Below are the features we consider and the corresponding intuition.

\begin{enumerate}[leftmargin=*]
  \item \textbf{The number of tokens in an example input}. Inputs with more tokens could contain a higher amount of information, allowing the attacker to more easily make correct inferences on the membership of the example. We use white space to split the input into tokens and count the number of tokens.
  \item \textbf{The number of tokens in the model output}. Longer outputs can potentially reveal more information about the model, which might make the model more susceptible to attacks. Similar to the input, we use white space to split the output into tokens and count the number of output tokens.
  \item \textbf{Victim model's perplexity on the example}. The perplexity of the victim model on an example can be used to measure the model's confidence in its prediction. A low perplexity score typically indicates that the language model has learned the patterns in the training data that are relevant to the example, suggesting that the example might be part of the training data. We follow the definition of perplexity in Section~\ref{subsubsec:metrics} to compute the perplexity of the victim model on each example.
  \item \textbf{The edit distance between the victim model output and the ground truth}. The edit distance measures the minimum number of operations (insertion, deletion, or substitution) required to transform one string (the model output) into another (the ground truth). A smaller edit distance indicates a closer match between the model output and the ground truth, suggesting that the model performs well on the input, meaning that this input is more likely to be in the training data. We use a python package called \texttt{nltk} to compute the edit distance between the model output and the ground truth.
  \item \textbf{The number of variable names in the example}. Source code contains shared keywords (e.g., \texttt{while}, \texttt{def}, etc), which are common to all programs written in a particular programming language, and human-defined variables, which are unique to specific programs. Intuitively, The presence of unique variable names in the code can make an example more identifiable and distinguishable from other examples in the dataset. We use a python package called \texttt{tree-sitter}~\cite{tree-sitter} to extract the variable names from the input and count their number of occurrences.
  \item \textbf{The BLEU score between the model output and the ground truth}. The BLEU score is a metric used to evaluate the quality of machine-generated content. A higher BLEU score indicates that the model output is more similar to the ground truth, suggesting that the model may have been trained on this example. We use the \texttt{sentence-bleu} function in the \texttt{nltk} package to compute the BLEU score between the model output and the ground truth.
\end{enumerate}

We split the evaluation examples into two groups: examples that are successfully inferred by the attacker and examples that are not.
Then, for examples in each group, we compute the features described above.
Taking the perplexity as an example, we obtain two lists of perplexity scores for the two groups of examples.
We conduct a Wilcoxon rank-sum test~\cite{Wilcoxon} to determine whether the two groups of examples have statistically significant differences in the feature values.
The Wilcoxon rank-sum test is a non-parametric statistical hypothesis test used to compare two related samples, which is widely used in the literature to understand the feature differences between two groups of examples~\cite{fan2021makes,asrdebugger,yang2023users}.
The null hypothesis is that there is no significant difference between the two groups of examples in terms of the feature values.
If the $p$-value of the test is less than 0.05, we reject the null hypothesis and conclude that there is a significant difference between the two groups of examples in terms of the feature values.
The statistical testing results are presented in Table~\ref{tab:feature-diff}.
From the table, we observe that there are statistically significant differences (with $p$-values less than 0.01) between the successfully and unsuccessfully inferred examples for all the features analyzed.

In addition, following previous studies~\cite{fan2021makes,asrdebugger,yang2023users} we calculate the Cohen's effect sizes ($\delta$), which are statistical measures used to quantify the magnitude of differences or relationships.
To determine the significance of the effect size, we adopt a guideline~\cite{cohen1988statistical} which states that an effect size of $|\delta|$ less than 0.2 is considered negligible, between 0.2 and 0.5 is small, between 0.5 and 0.8 is medium, and larger than 0.8 is large.
The effect size of these differences varies across features.
For input length, output length, edit distance, and the number of variables, the effect size is negligible, indicating that while there are statistically significant differences between the two groups, these differences may not have a substantial impact on the susceptibility of the model to membership inference attacks.

The effect size for perplexity is larger than that of other features, suggesting that these features may have a more significant impact on the model's vulnerability to membership inference attacks.
A lower perplexity score for successfully attacked examples indicates that the victim model has a higher confidence in its predictions for these examples, potentially because the model has learned patterns in the training data relevant to these examples.
In summary, our analysis reveals that the differences in feature values between the two groups of examples are statistically significant. However, the effect size is small or negligible for most features, indicating that these features alone may not be strong indicators of a model's susceptibility to membership inference attacks.

\begin{table}[!t]
    \centering
    \caption{The difference between the feature values of successfully attacked and unsuccessfully attacked examples.}
    \begin{tabular}{lcccc}
    \toprule
    Features            & Success & Unsuccess & $p$-value          & Effect size   \\ \midrule
    Input length       & 326.07    & 285.33   & $<$0.01    & Negligible      \\
    Output length        & 6.06   & 5.67  & $<$0.01    & Negligible     \\
    Perplexity & 9.190     & 27.08    & $<$0.01    & Small      \\
    Edit Distance     & 16.21    & 18.83    & $<$0.01    & Negligible \\
    BLEU Score     & 16.21    & 18.83    & $<$0.01    & Negligible \\
    No. variables     & 6.66     & 5.81     & $<$0.01 & Negligible \\ \bottomrule
    \end{tabular}
    \label{tab:feature-diff}
\end{table}

\begin{tcolorbox}[boxrule=0pt,frame hidden,sharp corners,enhanced,borderline north={1pt}{0pt}{black},borderline south={1pt}{0pt}{black},boxsep=2pt,left=2pt,right=2pt,top=2.5pt,bottom=2pt]
  \textbf{Answers to RQ3}:   MIA classifiers tend to perform better on examples that have lower perplexity scores. However, input length, output length, edit distance, and the number of variables show negligible effect sizes.
\end{tcolorbox}

\section{Discussion}
\label{sec:discussion}

\subsection{How to defense against MIA?}

For domains that are well studied, e.g., image classification, researchers have proposed a series of defensive methods to protect models, e.g., DP-SGD~\cite{DP-SGD}, model ensemble~\cite{tang2022mitigating}, and adversarial regularization~\cite{Nasr-defense}.
However, to the best of our knowledge, there is no existing work designed for protecting code generation models.
By default, the CodeGPT model uses beam search~\cite{beam-search} as the decoding strategy.
We let the model use another decoding strategy called ``top-$k$ sampling''~\cite{bengio2015scheduled} to generate code.
This approach selects the $k$ most likely tokens from the probability distribution at each time step and then samples from those tokens to generate the next word in the sequence.
This method can result in more diverse and creative outputs compared to beam search, which tends to generate more focused and deterministic outputs.
By introducing more randomness and diversity into the generated outputs, top-$k$ sampling can make it harder for an attacker to correlate the outputs with specific training data points.

There are two important hyperparameters in the top-$k$ sampling strategy~\cite{bengio2015scheduled}: (1) the value of $k$ and (2) the \textit{temperature}.
The `$k$' value determines the number of most likely tokens to consider for sampling at each step.
A smaller `$k$' results in a more focused set of tokens, leading to more deterministic and coherent text generation, but may also cause the output to be repetitive and less creative.
A larger `$k$' allows for a more diverse set of tokens to be considered, resulting in more diverse and creative outputs.
The temperature is a scaling factor applied to the logits before converting them into probabilities using the softmax function.
A higher temperature value flattens the distribution, promoting diversity and creativity in the text but potentially leading to less coherent and more random outputs.
We change the decoding strategy from beam search to top-$k$ sampling.
We also try different combinations of the `$k$' and temperature to see whether the leakage risk can be mitigated.

Table~\ref{tab:decoding} illustrates how the top-$k$ sampling strategy and its hyperparameters can affect the risk of membership leakage.
We can observe that changing the decoding strategy can mitigate such risk.
To be more specific, the AUC score changes from 0.98 to 0.59, showing that the top-$k$ sampling strategy can reduce the risk of membership leakage by around 40\%.
However, the performance of MIA is not sensitive to the choice of values of $k$ and temperature.
We try 7 different combinations of the `$k$' value and the temperature, and the AUC score is always around 0.59.
This simple defense strategy has minimal impact on the performance of the victim model.
As indicated by the BLUE score, when the temperature value is low (e.g., less than 1), the BLUE score is close to that of the beam search decoding strategy (0.59).
The result is consistent with the finding from the ablation study in RQ1 that the model output contributes most to the effectiveness of the proposed approach.

\begin{table}[!t]
    \centering
    \caption{How victim model's performance and attack results changes when using different decoding hyperparameters.}
    \begin{tabular}{@{}lrrrrrr@{}}
    \toprule
    Variation         & $k$   & $temp$ & BLUE & Power & Error & AUC   \\ \midrule
    \multirow{4}{*}{Vary Temp}  & 50  & 0.1  & 0.63  & 0.64  & 0.52  & 0.59 \\
                                       & 50  & 0.5  & 0.61  & 0.64  & 0.52  & 0.59 \\
                                       & 50  & 1    & 0.57  & 0.64  & 0.52  & 0.59 \\
                                       & 50  & 2    & 0.36  & 0.64  & 0.52  & 0.59 \\ \cmidrule(lr){2-7}
    \multirow{3}{*}{Vary k}            & 10  & 1    & 0.58  & 0.64  & 0.52  & 0.59 \\
                                       & 50  & 1    & 0.57  & 0.64  & 0.52  & 0.59 \\
                                       & 100 & 1    & 0.57  & 0.64  & 0.52  & 0.59 \\ \cmidrule(lr){1-7}
    \multicolumn{3}{c}{Beam-search to decode}    & 0.59 & 0.95  & 0.10  & 0.98 \\ \bottomrule
    \end{tabular}
    \label{tab:decoding}
    \end{table}

\subsection{Ethical Considerations}
The progress in code models and their applications can greatly benefit society, but it is crucial to consider the potential privacy and security risks associated with them.
Our aim is not to promote or facilitate malicious behavior, but rather to raise awareness of the potential risks associated with code models and to contribute to the development of secure and privacy-preserving code models.

It is imperative that ethical considerations are taken into account when using code models and other machine learning models.
This includes responsible data handling and privacy protection, as well as ensuring that these models are used for the benefit of society.
As researchers and practitioners in the field of software engineering, it is our responsibility to ensure that our work is used for ethical and beneficial purposes.
We hope that this paper will contribute to the ongoing discussion on the ethical use of code models and other machine learning models, and help to promote responsible and ethical research practices.

\subsection{Threats to Validity}

\vspace{0.2cm}
\noindent \textbf{Threats to Internal Validity.}
While we have explored the risk of membership inference leakage against various MIA attacks, we acknowledge that the performance of these methods may be affected by the choice of hyperparameters and randomness.
To mitigate these threats, we repeat the experiments 3 times and report the average results.
We use the hyperparameters in the original paper~\cite{CodeXGLUE} to train the victim model and implement the attacks.
When evaluating the metric ranking-based methods, we set the cut-off point as 50\% to determine whether an example is in the training data because the evaluation dataset is balanced.
In practice, the attacker can adjust the cut-off position based on their specific objectives and the nature of the dataset they are targeting.
For example, if an attacker aims to increase the true positive rate, they may opt for a higher cut-off position, such as 70\% or 80\%. While this approach may increase the likelihood of including more actual training examples, it also comes with a trade-off: the false positive rate will likely rise, leading to more non-training examples being incorrectly classified as training examples.

We also acknowledge that the choice of experiment designs may affect the results.
For example, to evaluate the impact of each factor on the attack performance, we leave other factors unchanged, vary one factor, and observe the change in the attack performance.
Such evaluation settings that analyze each factor separately are also widely adopted in the literature that analyzes AI systems. For example, Yan et al.~\cite{FSE_Yan} analyze how training data size affects model robustness by fixing other hyperparameters and gradually increasing training data size to compare the robustness of each model, which is similar to our setting for analyzing how the number of training epochs affects privacy risks.
We plan to investigate the interaction between different factors and their impacts on privacy risks (e.g., using Design of Experiments~\cite{fisher1966design}) in future work.

\vspace{0.2cm}
\noindent \textbf{Threats to External Validity.} In the context of the paper, external validity is related to the generalizability of the findings regarding the membership leakage risk of CodeGPT to other code models.
The findings may not generalize to other code models or language models with different architectures, e.g., LSTM.
The paper focuses on the CodeGPT model, a recently proposed model that leverages the GPT architecture.
The GPT architecture is the foundation of many state-of-the-art language models, e.g., ChatGPT.
The selection of surrogate models may also affect the membership inference results. To understand this potential impact, we evaluate four different surrogate models in RQ2 and confirm that in all four settings our method outperforms the baselines.
This study uses the JavaCorpus dataset in the experiments, a large collection of Java code.
We acknowledge the results may not generalize to other programming languages, e.g., Python, C++, etc.
We believe that this threat is minimal as our method is programming language agnostic and can be extended to other programming languages.
We leave evaluation on other programming languages as future work and encourage other researchers to validate our findings on more datasets, including non-Java datasets.

Moreover, while our study uses a balanced dataset for evaluation, we acknowledge that in real-world scenarios, datasets often exhibit imbalanced class distributions. In such cases, attackers might need to adapt their strategies accordingly, potentially selecting cut-off points that better reflect the underlying distribution.

The victim models evaluated in this paper are static, i.e., they are fine-tuned once and do not change during the attack.
The results obtained in this paper may not generalize to dynamic models that are continuously updated or retrained.
Training a new membership information classifier should mitigate such potential threats.
The assumption of static victim models comes from previous studies and is widely adopted~\cite{hisamoto-etal-2020-membership,carlini21extracting,alert,you-see,9825895}.
The investigation of MIA for dynamically changing models deserves a separate line of work, and we leave it for future work.

\section{Related Work}
\label{sec:related_work}

\subsection{Code Models and Threats}
Recent studies have highlighted that code models are vulnerable to various attacks and threats.
Yefet et al.~\cite{Yefet2020} employed the Fast Gradient Sign Method~\cite{FGSM} to adversarially transform source code, resulting in changes to the output of code models such as code2vec, GGNN, and GNN-FiLM.
Yang et al.~\cite{alert} emphasized the naturalness requirement in creating adversarial examples of code.
Srikant et al.\cite{Epresentation2021} used stronger adversarial algorithms (PGD~\cite{PGD}) to generate adversarial examples with higher success rates. These studies utilize white-box information (e.g., model parameters, gradients) to conduct attacks, however, there are also works on black-box attacks where an attacker only has access to the model's input and output.
Zhang et al.~\cite{MHM} modeled code attacks as a stochastic process and designed the Metropolis-Hastings Modifier (MHM) to generate adversarial examples for code.
Wei et al.~\cite{9916170} proposed a coverage-guided fuzzing algorithm to test code models.
Additionally, several works have evaluated code models against adversarial attacks using semantic preserving transformations~\cite{rabin2021generalizability,9678706,9438605,codeattack}.

There are new threats emerging for code models.
Nguyen et al.~\cite{coffee} conducted data poisoning attacks on API recommendation systems, finding that all three investigated systems were vulnerable to attacks that simply injected small amounts of malicious data into the training set.
Schuster et al.~\cite{263874} performed data poisoning attacks on code completion models, showing that by injecting malicious code snippets into the training set, the code completion models produced code with security vulnerabilities (e.g., using insecure APIs when encryption) in critical contexts.
Data poisoning can also be used to inject backdoors into code models~\cite{codebackdoor}.
Wan et al.~\cite{you-see} used fixed and grammar triggers to implant backdoors in code search models. Yang et al.~\cite{advdoor} proposed the use of adversarial features to create stealthy backdoors in code models.
Li et al.~\cite{CodePoisoner} leveraged another code model to generate dynamic backdoors in code models.
Data poisoning can also be used as a protection mechanism.
Sun et al.~\cite{CoProtector} proposed using data poisoning to prevent open-source data from being trained without authorization.

To the best of our knowledge, our study presents the first systematic investigation of membership inference attacks on code models.\footnote{We put the preprint of this paper on arXiv on 2 October 2023. URL: \url{https://arxiv.org/abs/2310.01166}}
The existing works and this study demonstrate the vulnerability of code models, highlighting the need for vulnerability evaluation and mitigation techniques to protect against these types of attacks.

\subsection{Privacy Attacks on DNN Models}

This paper investigates the MIA~\cite{7958568,quantifying,truex2019demystifying}, which can serve as the gate to a series of other privacy attacks.
Data extraction attack~\cite{carlini21extracting} aims to extract training data from a victim model.
Model extraction attacks~\cite{krishna2019thieves,steal-rl} are designed to steal information about the victim model, e.g., model parameters, model architecture, model functionality, etc.
However, defense mechanisms for generation models (including text generation) have not been as extensively studied as those for other types of tasks, such as classification.
For example, Nasr et al.~\cite{Nasr-defense} leverage adversarial training as a defense against MIA.
Inspired by differential privacy, researchers also propose to train models with differential privacy~\cite{DP-SGD} to protect the models.
The proposed methods only demonstrate effectiveness on classification tasks.
Both adversarial training and differential privacy training are extremely expensive and may not be feasible for large-scale generative models like CodeGPT.

Privacy attacks can target at different types of data.
A large portion of effort has been devoted to the privacy attacks on image data~\cite{8429311,nasr2019comprehensive,truex2019demystifying,10.1145/3133956.3134012},
However, other types of data, such as tabular, text, and time-series data, have received comparatively less attention.
Popular benchmarks of tabular data in the context of privacy attacks include the UCI's diabetes dataset, German Credit Dataset~\cite{diabetes}, Adult Income Dataset~\cite{adult}, etc.
Some datasets of texts that contain sensitive information also suffer from privacy attacks, e.g., Yelp healthcare-related reviews~\cite{Tran2017Online}.
To the best of our knowledge, our paper presents the first study on privacy attacks on code models and datasets.
Privacy attacks can target at different types of tasks.
Classification tasks are the most prevalent type of tasks in privacy attacks, e.g., image classification~\cite{7958568}, income classification~\cite{10.5555/3241094.3241142}, text classification~\cite{quantifying}, etc.
Less attention has been paid to other types of tasks, such as generation tasks.
In generation tasks, the privacy risk of generative adversarial network (GAN) models is more well-studied~\cite{gan-leak}.
Another important task in generation tasks is the text generation.
Hisamoto et al.~\cite{hisamoto-etal-2020-membership} conduct MIA on machine translation systems.
Carlini et al.~\cite{carlini21extracting} leverage MIA to extract training data from text models.
This paper proposes an effective attack on code models and uses the two attacks as baselines.

Our study also suggests insights from the previous study~\cite{8429311} may not generalize to code models.
For example, Yeom et al.~\cite{8429311} show that the number of training epochs can affect the membership leakage risk while our experiment finds the effect is negligible.
This discrepancy may be due to the different characteristics of the models, data, and tasks, studied in the two papers.
Their conclusion is drawn from image classification and regression tasks on small models, while our study focuses on code completion tasks on large language models.

\section{Conclusion and Future Work}
\label{sec:conclusion}

In conclusion, this paper has shed light on the significant privacy concerns surrounding the use of code models, specifically in terms of membership information leakage.
We introduce \toolname, a novel membership inference attack method, and evaluated its efficacy against CodeGPT, an open-source code completion model.
Our findings reveal that the risk of membership inference attacks is alarmingly high, with \toolname achieving a high true positive rate 0.95 and a low false positive rate 0.10.
Furthermore, we demonstrate that an attacker's chances of success increase with more knowledge of the victim model, such as its architecture.
These findings serve as a call to action for the research community to pay greater attention to the privacy implications of code models and to develop more effective countermeasures against privacy attacks. Future work should aim to investigate more sophisticated defense mechanisms, explore other potential privacy risks, and establish best practices for the secure and responsible use of code models.

In future work, we plan to investigate membership leakage risks in code models with different architectures and programming languages.
Also, we plan to design more effective countermeasures against membership inference attacks.

\begin{tcolorbox}[colback=white, colframe=black]
    \textit{The replication package is available at \url{https://github.com/yangzhou6666/MIA-LLM4Code}, which is intended for academic and research purposes only.
    We do not condone or support the use of the replication package for malicious purposes.}
\end{tcolorbox}

\ifCLASSOPTIONcompsoc
  \section*{Acknowledgments}
\else
  \section*{Acknowledgment}
\fi

This research / project is supported by the National Research Foundation, under its Investigatorship Grant (NRF-NRFI08-2022-0002). Any opinions, findings and conclusions or recommendations expressed in this material are those of the author(s) and do not reflect the views of National Research Foundation, Singapore.

\balance
\bibliographystyle{IEEEtran}
\bibliography{reference}

\begin{thebibliography}{10}
\providecommand{\url}[1]{#1}
\csname url@samestyle\endcsname
\providecommand{\newblock}{\relax}
\providecommand{\bibinfo}[2]{#2}
\providecommand{\BIBentrySTDinterwordspacing}{\spaceskip=0pt\relax}
\providecommand{\BIBentryALTinterwordstretchfactor}{4}
\providecommand{\BIBentryALTinterwordspacing}{\spaceskip=\fontdimen2\font plus
\BIBentryALTinterwordstretchfactor\fontdimen3\font minus \fontdimen4\font\relax}
\providecommand{\BIBforeignlanguage}[2]{{%
\expandafter\ifx\csname l@#1\endcsname\relax
\typeout{** WARNING: IEEEtran.bst: No hyphenation pattern has been}%
\typeout{** loaded for the language `#1'. Using the pattern for}%
\typeout{** the default language instead.}%
\else
\language=\csname l@#1\endcsname
\fi
#2}}
\providecommand{\BIBdecl}{\relax}
\BIBdecl

\bibitem{bert}
\BIBentryALTinterwordspacing
J.~Devlin, M.-W. Chang, K.~Lee, and K.~Toutanova, ``{BERT}: Pre-training of deep bidirectional transformers for language understanding,'' in \emph{Proceedings of the 2019 Conference of the North {A}merican Chapter of the Association for Computational Linguistics: Human Language Technologies, Volume 1 (Long and Short Papers)}.\hskip 1em plus 0.5em minus 0.4em\relax Minneapolis, Minnesota: Association for Computational Linguistics, Jun. 2019, pp. 4171--4186. [Online]. Available: \url{https://aclanthology.org/N19-1423}
\BIBentrySTDinterwordspacing

\bibitem{RoBERTa}
\BIBentryALTinterwordspacing
Y.~Liu, M.~Ott, N.~Goyal, J.~Du, M.~Joshi, D.~Chen, O.~Levy, M.~Lewis, L.~Zettlemoyer, and V.~Stoyanov, ``Roberta: {A} robustly optimized {BERT} pretraining approach,'' \emph{CoRR}, vol. abs/1907.11692, 2019. [Online]. Available: \url{http://arxiv.org/abs/1907.11692}
\BIBentrySTDinterwordspacing

\bibitem{T5}
\BIBentryALTinterwordspacing
C.~Raffel, N.~Shazeer, A.~Roberts, K.~Lee, S.~Narang, M.~Matena, Y.~Zhou, W.~Li, and P.~J. Liu, ``Exploring the limits of transfer learning with a unified text-to-text transformer,'' \emph{Journal of Machine Learning Research}, vol.~21, no. 140, pp. 1--67, 2020. [Online]. Available: \url{http://jmlr.org/papers/v21/20-074.html}
\BIBentrySTDinterwordspacing

\bibitem{plbart}
W.~Ahmad, S.~Chakraborty, B.~Ray, and K.-W. Chang, ``Unified pre-training for program understanding and generation,'' in \emph{Proceedings of the 2021 Conference of the North American Chapter of the Association for Computational Linguistics: Human Language Technologies}.\hskip 1em plus 0.5em minus 0.4em\relax Online: Association for Computational Linguistics, Jun. 2021, pp. 2655--2668.

\bibitem{CodeXGLUE}
S.~Lu, D.~Guo, S.~Ren, J.~Huang, A.~Svyatkovskiy, A.~Blanco, C.~B. Clement, D.~Drain, D.~Jiang, D.~Tang, G.~Li, L.~Zhou, L.~Shou, L.~Zhou, M.~Tufano, M.~Gong, M.~Zhou, N.~Duan, N.~Sundaresan, S.~K. Deng, S.~Fu, and S.~Liu, ``Codexglue: {A} machine learning benchmark dataset for code understanding and generation,'' \emph{CoRR}, 2021.

\bibitem{husain2019codesearchnet}
H.~Husain, H.-H. Wu, T.~Gazit, M.~Allamanis, and M.~Brockschmidt, ``{CodeSearchNet} challenge: Evaluating the state of semantic code search,'' \emph{arXiv preprint arXiv:1909.09436}, 2019.

\bibitem{codellm_survey}
X.~Hou, Y.~Zhao, Y.~Liu, Z.~Yang, K.~Wang, L.~Li, X.~Luo, D.~Lo, J.~Grundy, and H.~Wang, ``Large language models for software engineering: A systematic literature review,'' 2023.

\bibitem{9794048}
M.~Izadi, R.~Gismondi, and G.~Gousios, ``Codefill: Multi-token code completion by jointly learning from structure and naming sequences,'' in \emph{2022 IEEE/ACM 44th International Conference on Software Engineering (ICSE)}, 2022, pp. 401--412.

\bibitem{9825884}
\BIBentryALTinterwordspacing
C.~Yang, B.~Xu, J.~Khan, G.~Uddin, D.~Han, Z.~Yang, and D.~Lo, ``Aspect-based api review classification: How far can pre-trained transformer model go?'' in \emph{2022 IEEE International Conference on Software Analysis, Evolution and Reengineering (SANER)}.\hskip 1em plus 0.5em minus 0.4em\relax Los Alamitos, CA, USA: IEEE Computer Society, mar 2022, pp. 385--395. [Online]. Available: \url{https://doi.ieeecomputersociety.org/10.1109/SANER53432.2022.00054}
\BIBentrySTDinterwordspacing

\bibitem{TechSumBot}
\BIBentryALTinterwordspacing
C.~Yang, B.~Xu, F.~Thung, Y.~Shi, T.~Zhang, Z.~Yang, X.~Zhou, J.~Shi, J.~He, D.~Han, and D.~Lo, \emph{Answer Summarization for Technical Queries: Benchmark and New Approach}.\hskip 1em plus 0.5em minus 0.4em\relax New York, NY, USA: Association for Computing Machinery, 2023. [Online]. Available: \url{https://doi.org/10.1145/3551349.3560421}
\BIBentrySTDinterwordspacing

\bibitem{9462962}
A.~Mazuera-Rozo, A.~Mojica-Hanke, M.~Linares-Vásquez, and G.~Bavota, ``Shallow or deep? an empirical study on detecting vulnerabilities using deep learning,'' in \emph{2021 IEEE/ACM 29th International Conference on Program Comprehension (ICPC)}, 2021, pp. 276--287.

\bibitem{codex}
M.~Chen, J.~Tworek, and H.~J. et~al., ``Evaluating large language models trained on code,'' \emph{CoRR}, 2021.

\bibitem{alert}
\BIBentryALTinterwordspacing
Z.~Yang, J.~Shi, J.~He, and D.~Lo, ``Natural attack for pre-trained models of code,'' in \emph{Proceedings of the 44th International Conference on Software Engineering}, ser. ICSE '22.\hskip 1em plus 0.5em minus 0.4em\relax New York, NY, USA: Association for Computing Machinery, 2022, p. 1482–1493. [Online]. Available: \url{https://doi.org/10.1145/3510003.3510146}
\BIBentrySTDinterwordspacing

\bibitem{Yefet2020}
N.~Yefet, U.~Alon, and E.~Yahav, ``Adversarial examples for models of code,'' \emph{Proc. {ACM} Program. Lang.}, vol.~4, no. {OOPSLA}, pp. 162:1--162:30, 2020.

\bibitem{Epresentation2021}
S.~Srikant, S.~Liu, T.~Mitrovska, S.~Chang, Q.~Fan, G.~Zhang, and U.~O'Reilly, ``{Generating Adversarial Computer Programs using Optimized Obfuscations},'' \emph{ICLR}, vol.~16, pp. 209--226, 2021.

\bibitem{9825895}
J.~Henkel, G.~Ramakrishnan, Z.~Wang, A.~Albarghouthi, S.~Jha, and T.~Reps, ``Semantic robustness of models of source code,'' in \emph{2022 IEEE International Conference on Software Analysis, Evolution and Reengineering (SANER)}, 2022, pp. 526--537.

\bibitem{nguyen2023adversarial}
T.-D. Nguyen, Z.~Yang, X.~B.~D. Le, Patanamon, Thongtanunam, and D.~Lo, ``Adversarial attacks on code models with discriminative graph patterns,'' 2023.

\bibitem{263874}
R.~Schuster, C.~Song, E.~Tromer, and V.~Shmatikov, ``You autocomplete me: Poisoning vulnerabilities in neural code completion,'' in \emph{30th USENIX Security Symposium (USENIX Security 21)}.\hskip 1em plus 0.5em minus 0.4em\relax USENIX Association, Aug. 2021, pp. 1559--1575.

\bibitem{you-see}
\BIBentryALTinterwordspacing
Y.~Wan, S.~Zhang, H.~Zhang, Y.~Sui, G.~Xu, D.~Yao, H.~Jin, and L.~Sun, ``You see what i want you to see: Poisoning vulnerabilities in neural code search,'' in \emph{Proceedings of the 30th ACM Joint European Software Engineering Conference and Symposium on the Foundations of Software Engineering}, ser. ESEC/FSE 2022.\hskip 1em plus 0.5em minus 0.4em\relax New York, NY, USA: Association for Computing Machinery, 2022, p. 1233–1245. [Online]. Available: \url{https://doi.org/10.1145/3540250.3549153}
\BIBentrySTDinterwordspacing

\bibitem{coffee}
P.~T. Nguyen, C.~Di~Sipio, J.~Di~Rocco, M.~Di~Penta, and D.~Di~Ruscio, ``Adversarial attacks to api recommender systems: Time to wake up and smell the coffee?'' in \emph{2021 36th IEEE/ACM International Conference on Automated Software Engineering (ASE)}, 2021, pp. 253--265.

\bibitem{yang2023memorzation}
\BIBentryALTinterwordspacing
Z.~Yang, Z.~Zhao, C.~Wang, J.~Shi, D.~Kim, D.~Han, and D.~Lo, ``Unveiling memorization in code models,'' in \emph{Proceedings of the IEEE/ACM 46th International Conference on Software Engineering}, ser. ICSE '24.\hskip 1em plus 0.5em minus 0.4em\relax New York, NY, USA: Association for Computing Machinery, 2024. [Online]. Available: \url{https://doi.org/10.1145/3597503.3639074}
\BIBentrySTDinterwordspacing

\bibitem{291327}
L.~Niu, S.~Mirza, Z.~Maradni, and C.~P{\"o}pper, ``{CodexLeaks}: Privacy leaks from code generation language models in {GitHub} copilot,'' in \emph{32nd USENIX Security Symposium (USENIX Security 23)}.\hskip 1em plus 0.5em minus 0.4em\relax Anaheim, CA: USENIX Association, Aug. 2023, pp. 2133--2150.

\bibitem{10006873}
M.~L. Siddiq, S.~H. Majumder, M.~R. Mim, S.~Jajodia, and J.~C.~S. Santos, ``An empirical study of code smells in transformer-based code generation techniques,'' in \emph{2022 IEEE 22nd International Working Conference on Source Code Analysis and Manipulation (SCAM)}, 2022, pp. 71--82.

\bibitem{CoProtector}
Z.~Sun, X.~Du, F.~Song, M.~Ni, and L.~Li, ``Coprotector: Protect open-source code against unauthorized training usage with data poisoning,'' in \emph{Proceedings of the ACM Web Conference 2022}, ser. WWW '22.\hskip 1em plus 0.5em minus 0.4em\relax New York, NY, USA: Association for Computing Machinery, 2022, p. 652–660.

\bibitem{advdoor}
Z.~Yang, B.~Xu, J.~M. Zhang, H.~J. Kang, J.~Shi, J.~He, and D.~Lo, ``Stealthy backdoor attack for code models,'' \emph{IEEE Transactions on Software Engineering}, no.~01, pp. 1--21, feb.

\bibitem{AWSCodeWhisperer}
``Aws codewhisperer: Features,'' \url{https://aws.amazon.com/codewhisperer/features/}, accessed: March 29, 2023.

\bibitem{basak2023secretbench}
S.~K. Basak, L.~Neil, B.~Reaves, and L.~Williams, ``Secretbench: A dataset of software secrets,'' in \emph{Proceedings of the 20th International Conference on Mining Software Repositories}, ser. MSR '23, 2023.

\bibitem{codegen}
E.~Nijkamp, B.~Pang, H.~Hayashi, L.~Tu, H.~Wang, Y.~Zhou, S.~Savarese, and C.~Xiong, ``Codegen: An open large language model for code with multi-turn program synthesis,'' in \emph{The Eleventh International Conference on Learning Representations}, 2023.

\bibitem{codeparrot}
\BIBentryALTinterwordspacing
``codeparrot (codeparrot),'' huggingface.co. [Online]. Available: \url{https://huggingface.co/codeparrot}
\BIBentrySTDinterwordspacing

\bibitem{gpt-neo}
\BIBentryALTinterwordspacing
S.~Black, L.~Gao, P.~Wang, C.~Leahy, and S.~Biderman, ``{GPT-Neo: Large Scale Autoregressive Language Modeling with Mesh-Tensorflow},'' Mar. 2021, {If you use this software, please cite it using these metadata.} [Online]. Available: \url{https://doi.org/10.5281/zenodo.5297715}
\BIBentrySTDinterwordspacing

\bibitem{10.1145/3520312.3534862}
\BIBentryALTinterwordspacing
F.~F. Xu, U.~Alon, G.~Neubig, and V.~J. Hellendoorn, ``A systematic evaluation of large language models of code,'' in \emph{Proceedings of the 6th ACM SIGPLAN International Symposium on Machine Programming}, ser. MAPS 2022.\hskip 1em plus 0.5em minus 0.4em\relax New York, NY, USA: Association for Computing Machinery, 2022, p. 1–10. [Online]. Available: \url{https://doi.org/10.1145/3520312.3534862}
\BIBentrySTDinterwordspacing

\bibitem{javacorpus}
\BIBentryALTinterwordspacing
M.~Allamanis and C.~Sutton, ``Mining source code repositories at massive scale using language modeling,'' in \emph{2013 10th IEEE Working Conference on Mining Software Repositories (MSR 2013)}.\hskip 1em plus 0.5em minus 0.4em\relax Los Alamitos, CA, USA: IEEE Computer Society, may 2013, pp. 207--216. [Online]. Available: \url{https://doi.ieeecomputersociety.org/10.1109/MSR.2013.6624029}
\BIBentrySTDinterwordspacing

\bibitem{hisamoto-etal-2020-membership}
\BIBentryALTinterwordspacing
S.~Hisamoto, M.~Post, and K.~Duh, ``Membership inference attacks on sequence-to-sequence models: {I}s my data in your machine translation system?'' \emph{Transactions of the Association for Computational Linguistics}, vol.~8, pp. 49--63, 2020. [Online]. Available: \url{https://aclanthology.org/2020.tacl-1.4}
\BIBentrySTDinterwordspacing

\bibitem{carlini21extracting}
N.~Carlini, F.~Tramer, E.~Wallace, M.~Jagielski, A.~Herbert-Voss, K.~Lee, A.~Roberts, T.~Brown, D.~Song, U.~Erlingsson, A.~Oprea, and C.~Raffel, ``Extracting training data from large language models,'' in \emph{USENIX Security Symposium}, 2021.

\bibitem{beam-search}
\BIBentryALTinterwordspacing
M.~Freitag and Y.~Al-Onaizan, ``Beam search strategies for neural machine translation,'' in \emph{Proceedings of the First Workshop on Neural Machine Translation}.\hskip 1em plus 0.5em minus 0.4em\relax Vancouver: Association for Computational Linguistics, Aug. 2017, pp. 56--60. [Online]. Available: \url{https://aclanthology.org/W17-3207}
\BIBentrySTDinterwordspacing

\bibitem{CodeBERT}
Z.~Feng, D.~Guo, D.~Tang, N.~Duan, X.~Feng, M.~Gong, L.~Shou, B.~Qin, T.~Liu, D.~Jiang, and M.~Zhou, ``{C}ode{BERT}: A pre-trained model for programming and natural languages,'' in \emph{Findings of the Association for Computational Linguistics: EMNLP 2020}.\hskip 1em plus 0.5em minus 0.4em\relax Association for Computational Linguistics, Nov. 2020, pp. 1536--1547.

\bibitem{GraphCodeBERT}
D.~Guo, S.~Ren, S.~Lu, Z.~Feng, D.~Tang, S.~Liu, L.~Zhou, N.~Duan, A.~Svyatkovskiy, S.~F. andz Michele~Tufano, S.~K. Deng, C.~B. Clement, D.~Drain, N.~Sundaresan, J.~Yin, D.~Jiang, and M.~Zhou, ``Graphcodebert: Pre-training code representations with data flow,'' in \emph{9th International Conference on Learning Representations, {ICLR} 2021, Virtual Event, Austria, May 3-7, 2021}, 2021.

\bibitem{wang2021codet5}
Y.~Wang, W.~Wang, S.~Joty, and S.~C. Hoi, ``Codet5: Identifier-aware unified pre-trained encoder-decoder models for code understanding and generation,'' in \emph{Proceedings of the 2021 Conference on Empirical Methods in Natural Language Processing, EMNLP 2021}, 2021.

\bibitem{py150}
\BIBentryALTinterwordspacing
V.~Raychev, P.~Bielik, and M.~Vechev, ``Probabilistic model for code with decision trees,'' in \emph{Proceedings of the 2016 ACM SIGPLAN International Conference on Object-Oriented Programming, Systems, Languages, and Applications}, ser. OOPSLA 2016.\hskip 1em plus 0.5em minus 0.4em\relax New York, NY, USA: Association for Computing Machinery, 2016, p. 731–747. [Online]. Available: \url{https://doi.org/10.1145/2983990.2984041}
\BIBentrySTDinterwordspacing

\bibitem{10.1145/3510003.3510222}
\BIBentryALTinterwordspacing
H.~Ye, M.~Martinez, and M.~Monperrus, ``Neural program repair with execution-based backpropagation,'' in \emph{Proceedings of the 44th International Conference on Software Engineering}, ser. ICSE '22.\hskip 1em plus 0.5em minus 0.4em\relax New York, NY, USA: Association for Computing Machinery, 2022, p. 1506–1518. [Online]. Available: \url{https://doi.org/10.1145/3510003.3510222}
\BIBentrySTDinterwordspacing

\bibitem{7958568}
R.~Shokri, M.~Stronati, C.~Song, and V.~Shmatikov, ``Membership inference attacks against machine learning models,'' in \emph{2017 IEEE Symposium on Security and Privacy (SP)}, 2017, pp. 3--18.

\bibitem{zheng-jiang-2022-empirical}
\BIBentryALTinterwordspacing
X.~Zheng and J.~Jiang, ``An empirical study of memorization in {NLP},'' in \emph{Proceedings of the 60th Annual Meeting of the Association for Computational Linguistics (Volume 1: Long Papers)}.\hskip 1em plus 0.5em minus 0.4em\relax Dublin, Ireland: Association for Computational Linguistics, May 2022, pp. 6265--6278. [Online]. Available: \url{https://aclanthology.org/2022.acl-long.434}
\BIBentrySTDinterwordspacing

\bibitem{DBLP:conf/sp/PearceA0DK22}
\BIBentryALTinterwordspacing
H.~Pearce, B.~Ahmad, B.~Tan, B.~Dolan{-}Gavitt, and R.~Karri, ``Asleep at the keyboard? assessing the security of github copilot's code contributions,'' in \emph{43rd {IEEE} Symposium on Security and Privacy, {SP} 2022, San Francisco, CA, USA, May 22-26, 2022}.\hskip 1em plus 0.5em minus 0.4em\relax {IEEE}, 2022, pp. 754--768. [Online]. Available: \url{https://doi.org/10.1109/SP46214.2022.9833571}
\BIBentrySTDinterwordspacing

\bibitem{feitelson2021we}
D.~G. Feitelson, ``" we do not appreciate being experimented on": Developer and researcher views on the ethics of experiments on open-source projects,'' \emph{arXiv preprint arXiv:2112.13217}, 2021.

\bibitem{MHM}
H.~Zhang, Z.~Li, G.~Li, L.~Ma, Y.~Liu, and Z.~Jin, ``Generating adversarial examples for holding robustness of source code processing models,'' \emph{Proceedings of the AAAI Conference on Artificial Intelligence}, vol.~34, no.~01, pp. 1169--1176, Apr. 2020.

\bibitem{9609166}
X.~Zhou, D.~Han, and D.~Lo, ``Assessing generalizability of codebert,'' in \emph{2021 IEEE International Conference on Software Maintenance and Evolution (ICSME)}, 2021, pp. 425--436.

\bibitem{10.1145/3533767.3534390}
\BIBentryALTinterwordspacing
Z.~Zeng, H.~Tan, H.~Zhang, J.~Li, Y.~Zhang, and L.~Zhang, ``An extensive study on pre-trained models for program understanding and generation,'' in \emph{Proceedings of the 31st ACM SIGSOFT International Symposium on Software Testing and Analysis}, ser. ISSTA 2022.\hskip 1em plus 0.5em minus 0.4em\relax New York, NY, USA: Association for Computing Machinery, 2022, p. 39–51. [Online]. Available: \url{https://doi.org/10.1145/3533767.3534390}
\BIBentrySTDinterwordspacing

\bibitem{PTM4TAG}
\BIBentryALTinterwordspacing
J.~He, B.~Xu, Z.~Yang, D.~Han, C.~Yang, and D.~Lo, ``Ptm4tag: Sharpening tag recommendation of stack overflow posts with pre-trained models,'' ser. ICPC '22.\hskip 1em plus 0.5em minus 0.4em\relax New York, NY, USA: Association for Computing Machinery, 2022, p. 1–11. [Online]. Available: \url{https://doi.org/10.1145/3524610.3527897}
\BIBentrySTDinterwordspacing

\bibitem{he2023representation}
J.~He, Z.~Xin, B.~Xu, T.~Zhang, K.~Kim, Z.~Yang, F.~Thung, I.~Irsan, and D.~Lo, ``Representation learning for stack overflow posts: How far are we?'' 2023.

\bibitem{gpt-2}
A.~Radford, J.~Wu, R.~Child, D.~Luan, D.~Amodei, and I.~Sutskever, ``Language models are unsupervised multitask learners,'' 2019.

\bibitem{stupidbug}
\BIBentryALTinterwordspacing
K.~Jesse, T.~Ahmed, P.~T. Devanbu, and E.~Morgan, ``Large language models and simple, stupid bugs,'' in \emph{2023 IEEE/ACM 20th International Conference on Mining Software Repositories (MSR)}.\hskip 1em plus 0.5em minus 0.4em\relax Los Alamitos, CA, USA: IEEE Computer Society, may 2023, pp. 563--575. [Online]. Available: \url{https://doi.ieeecomputersociety.org/10.1109/MSR59073.2023.00082}
\BIBentrySTDinterwordspacing

\bibitem{big_code}
\BIBentryALTinterwordspacing
R.~Karampatsis, H.~Babii, R.~Robbes, C.~Sutton, and A.~Janes, ``Big code != big vocabulary: Open-vocabulary models for source code,'' in \emph{2020 IEEE/ACM 42nd International Conference on Software Engineering (ICSE)}.\hskip 1em plus 0.5em minus 0.4em\relax Los Alamitos, CA, USA: IEEE Computer Society, oct 2020, pp. 1073--1085. [Online]. Available: \url{https://doi.ieeecomputersociety.org/}
\BIBentrySTDinterwordspacing

\bibitem{10.3115/1218955.1219032}
\BIBentryALTinterwordspacing
C.-Y. Lin and F.~J. Och, ``Automatic evaluation of machine translation quality using longest common subsequence and skip-bigram statistics,'' in \emph{Proceedings of the 42nd Annual Meeting on Association for Computational Linguistics}, ser. ACL '04.\hskip 1em plus 0.5em minus 0.4em\relax USA: Association for Computational Linguistics, 2004, p. 605–es. [Online]. Available: \url{https://doi.org/10.3115/1218955.1219032}
\BIBentrySTDinterwordspacing

\bibitem{ppl}
\BIBentryALTinterwordspacing
S.~F. Chen and J.~Goodman, ``An empirical study of smoothing techniques for language modeling,'' \emph{Computer Speech and Language}, vol.~13, no.~4, pp. 359--394, 1999. [Online]. Available: \url{https://www.sciencedirect.com/science/article/pii/S0885230899901286}
\BIBentrySTDinterwordspacing

\bibitem{zlibnet}
``A massively spiffy yet delicately unobtrusive compression library,'' \url{https://zlib.net/}, accessed on March 27, 2023.

\bibitem{8429311}
S.~Yeom, I.~Giacomelli, M.~Fredrikson, and S.~Jha, ``Privacy risk in machine learning: Analyzing the connection to overfitting,'' in \emph{2018 IEEE 31st Computer Security Foundations Symposium (CSF)}, 2018, pp. 268--282.

\bibitem{montgomery2009design}
\BIBentryALTinterwordspacing
D.~Montgomery, \emph{Design and Analysis of Experiments 7th Edition with Student Solutions Manual and Design Expert 7. 0. 3 Set}.\hskip 1em plus 0.5em minus 0.4em\relax John Wiley \& Sons Canada, Limited, 2009. [Online]. Available: \url{https://books.google.com.sg/books?id=S2CQuAAACAAJ}
\BIBentrySTDinterwordspacing

\bibitem{cotroneo2023vulnerabilities}
D.~Cotroneo, C.~Improta, P.~Liguori, and R.~Natella, ``Vulnerabilities in ai code generators: Exploring targeted data poisoning attacks,'' 2023.

\bibitem{st1989analysis}
L.~St, S.~Wold \emph{et~al.}, ``Analysis of variance (anova),'' \emph{Chemometrics and intelligent laboratory systems}, vol.~6, no.~4, pp. 259--272, 1989.

\bibitem{Wilcoxon}
\BIBentryALTinterwordspacing
F.~Wilcoxon, ``Individual comparisons by ranking methods,'' \emph{Biometrics Bulletin}, vol.~1, no.~6, pp. 80--83, 1945. [Online]. Available: \url{http://www.jstor.org/stable/3001968}
\BIBentrySTDinterwordspacing

\bibitem{toothaker1993multiple}
L.~E. Toothaker, \emph{Multiple comparison procedures}.\hskip 1em plus 0.5em minus 0.4em\relax Sage, 1993, no.~89.

\bibitem{sedgwick2012multiple}
P.~Sedgwick, ``Multiple significance tests: the bonferroni correction,'' \emph{Bmj}, vol. 344, 2012.

\bibitem{tree-sitter}
T.~T. sitter Contributors, ``{Tree-sitter}: {A}n {A} incremental parsing library,'' \url{https://tree-sitter.github.io/tree-sitter/}, Accessed on 25th March, 2023.

\bibitem{fan2021makes}
Y.~Fan, X.~Xia, D.~Lo, A.~E. Hassan, and S.~Li, ``What makes a popular academic ai repository?'' \emph{Empirical Software Engineering}, vol.~26, no.~1, pp. 1--35, 2021.

\bibitem{asrdebugger}
\BIBentryALTinterwordspacing
Z.~Yang, J.~Shi, M.~H. Asyrofi, B.~Xu, X.~Zhou, D.~Han, and D.~Lo, ``Prioritizing speech test cases,'' 2023. [Online]. Available: \url{https://arxiv.org/abs/2302.00330}
\BIBentrySTDinterwordspacing

\bibitem{yang2023users}
Z.~Yang, C.~Wang, J.~Shi, T.~Hoang, P.~Kochhar, Q.~Lu, Z.~Xing, and D.~Lo, ``What do users ask in open-source ai repositories? an empirical study of github issues,'' in \emph{Proceedings of the 20th International Conference on Mining Software Repositories}, ser. MSR '23, 2023.

\bibitem{cohen1988statistical}
J.~Cohen, \emph{Statistical Power Analysis for the Behavioral Sciences}, 2nd~ed.\hskip 1em plus 0.5em minus 0.4em\relax Hillsdale, NJ: Lawrence Erlbaum Associates, 1988.

\bibitem{DP-SGD}
\BIBentryALTinterwordspacing
M.~Abadi, A.~Chu, I.~Goodfellow, H.~B. McMahan, I.~Mironov, K.~Talwar, and L.~Zhang, ``Deep learning with differential privacy,'' in \emph{Proceedings of the 2016 ACM SIGSAC Conference on Computer and Communications Security}, ser. CCS '16.\hskip 1em plus 0.5em minus 0.4em\relax New York, NY, USA: Association for Computing Machinery, 2016, p. 308–318. [Online]. Available: \url{https://doi.org/10.1145/2976749.2978318}
\BIBentrySTDinterwordspacing

\bibitem{tang2022mitigating}
X.~Tang, S.~Mahloujifar, L.~Song, V.~Shejwalkar, M.~Nasr, A.~Houmansadr, and P.~Mittal, ``Mitigating membership inference attacks by $\{$Self-Distillation$\}$ through a novel ensemble architecture,'' in \emph{31st USENIX Security Symposium (USENIX Security 22)}, 2022, pp. 1433--1450.

\bibitem{Nasr-defense}
\BIBentryALTinterwordspacing
M.~Nasr, R.~Shokri, and A.~Houmansadr, ``Machine learning with membership privacy using adversarial regularization,'' in \emph{Proceedings of the 2018 ACM SIGSAC Conference on Computer and Communications Security}, ser. CCS '18.\hskip 1em plus 0.5em minus 0.4em\relax New York, NY, USA: Association for Computing Machinery, 2018, p. 634–646. [Online]. Available: \url{https://doi.org/10.1145/3243734.3243855}
\BIBentrySTDinterwordspacing

\bibitem{bengio2015scheduled}
S.~Bengio, O.~Vinyals, N.~Jaitly, and N.~Shazeer, ``Scheduled sampling for sequence prediction with recurrent neural networks,'' \emph{Advances in neural information processing systems}, vol.~28, 2015.

\bibitem{FSE_Yan}
\BIBentryALTinterwordspacing
S.~Yan, G.~Tao, X.~Liu, J.~Zhai, S.~Ma, L.~Xu, and X.~Zhang, ``Correlations between deep neural network model coverage criteria and model quality,'' in \emph{Proceedings of the 28th ACM Joint Meeting on European Software Engineering Conference and Symposium on the Foundations of Software Engineering}, ser. ESEC/FSE 2020.\hskip 1em plus 0.5em minus 0.4em\relax New York, NY, USA: Association for Computing Machinery, 2020, p. 775–787. [Online]. Available: \url{https://doi.org/10.1145/3368089.3409671}
\BIBentrySTDinterwordspacing

\bibitem{fisher1966design}
R.~A. Fisher, R.~A. Fisher, S.~Genetiker, R.~A. Fisher, S.~Genetician, G.~Britain, R.~A. Fisher, and S.~G{\'e}n{\'e}ticien, \emph{The design of experiments}.\hskip 1em plus 0.5em minus 0.4em\relax Springer, 1966, vol.~21.

\bibitem{FGSM}
\BIBentryALTinterwordspacing
I.~Goodfellow, J.~Shlens, and C.~Szegedy, ``Explaining and harnessing adversarial examples,'' in \emph{International Conference on Learning Representations}, 2015. [Online]. Available: \url{http://arxiv.org/abs/1412.6572}
\BIBentrySTDinterwordspacing

\bibitem{PGD}
A.~Madry, A.~Makelov, L.~Schmidt, D.~Tsipras, and A.~Vladu, ``Towards deep learning models resistant to adversarial attacks,'' in \emph{6th International Conference on Learning Representations, {ICLR} 2018, Vancouver, BC, Canada, April 30 - May 3, 2018, Conference Track Proceedings}, 2018.

\bibitem{9916170}
M.~Wei, Y.~Huang, J.~Yang, J.~Wang, and S.~Wang, ``Cocofuzzing: Testing neural code models with coverage-guided fuzzing,'' \emph{IEEE Transactions on Reliability}, pp. 1--14, 2022.

\bibitem{rabin2021generalizability}
M.~R.~I. Rabin, N.~D. Bui, K.~Wang, Y.~Yu, L.~Jiang, and M.~A. Alipour, ``On the generalizability of neural program models with respect to semantic-preserving program transformations,'' \emph{Information and Software Technology}, vol. 135, p. 106552, 2021.

\bibitem{9678706}
L.~Applis, A.~Panichella, and A.~van Deursen, ``Assessing robustness of ml-based program analysis tools using metamorphic program transformations,'' in \emph{ASE 2021}, 2021, pp. 1377--1381.

\bibitem{9438605}
M.~V. Pour, Z.~Li, L.~Ma, and H.~Hemmati, ``A search-based testing framework for deep neural networks of source code embedding,'' in \emph{14th {IEEE} Conference on Software Testing, Verification and Validation, {ICST} 2021, Porto de Galinhas, Brazil, April 12-16, 2021}.\hskip 1em plus 0.5em minus 0.4em\relax {IEEE}, 2021.

\bibitem{codeattack}
\BIBentryALTinterwordspacing
A.~Jha and C.~K. Reddy, ``Codeattack: code-based adversarial attacks for pre-trained programming language models,'' in \emph{Proceedings of the Thirty-Seventh AAAI Conference on Artificial Intelligence and Thirty-Fifth Conference on Innovative Applications of Artificial Intelligence and Thirteenth Symposium on Educational Advances in Artificial Intelligence}, ser. AAAI'23/IAAI'23/EAAI'23.\hskip 1em plus 0.5em minus 0.4em\relax AAAI Press, 2023. [Online]. Available: \url{https://doi.org/10.1609/aaai.v37i12.26739}
\BIBentrySTDinterwordspacing

\bibitem{codebackdoor}
\BIBentryALTinterwordspacing
G.~Ramakrishnan and A.~Albarghouthi, ``Backdoors in neural models of source code,'' in \emph{2022 26th International Conference on Pattern Recognition (ICPR)}.\hskip 1em plus 0.5em minus 0.4em\relax Los Alamitos, CA, USA: IEEE Computer Society, aug 2022, pp. 2892--2899. [Online]. Available: \url{https://doi.ieeecomputersociety.org/10.1109/ICPR56361.2022.9956690}
\BIBentrySTDinterwordspacing

\bibitem{CodePoisoner}
\BIBentryALTinterwordspacing
J.~Li, Z.~Li, H.~Zhang, G.~Li, Z.~Jin, X.~Hu, and X.~Xia, ``Poison attack and defense on deep source code processing models,'' 2022. [Online]. Available: \url{https://arxiv.org/abs/2210.17029}
\BIBentrySTDinterwordspacing

\bibitem{quantifying}
\BIBentryALTinterwordspacing
F.~Mireshghallah, K.~Goyal, A.~Uniyal, T.~Berg-Kirkpatrick, and R.~Shokri, ``Quantifying privacy risks of masked language models using membership inference attacks,'' in \emph{Proceedings of the 2022 Conference on Empirical Methods in Natural Language Processing}.\hskip 1em plus 0.5em minus 0.4em\relax Abu Dhabi, United Arab Emirates: Association for Computational Linguistics, Dec. 2022, pp. 8332--8347. [Online]. Available: \url{https://aclanthology.org/2022.emnlp-main.570}
\BIBentrySTDinterwordspacing

\bibitem{truex2019demystifying}
S.~Truex, L.~Liu, M.~E. Gursoy, L.~Yu, and W.~Wei, ``Demystifying membership inference attacks in machine learning as a service,'' \emph{IEEE Transactions on Services Computing}, vol.~14, no.~6, pp. 2073--2089, 2019.

\bibitem{krishna2019thieves}
K.~Krishna, G.~S. Tomar, A.~P. Parikh, N.~Papernot, and M.~Iyyer, ``Thieves on sesame street! model extraction of bert-based apis,'' \emph{arXiv preprint arXiv:1910.12366}, 2019.

\bibitem{steal-rl}
\BIBentryALTinterwordspacing
K.~Chen, S.~Guo, T.~Zhang, X.~Xie, and Y.~Liu, ``Stealing deep reinforcement learning models for fun and profit,'' in \emph{Proceedings of the 2021 ACM Asia Conference on Computer and Communications Security}, ser. ASIA CCS '21.\hskip 1em plus 0.5em minus 0.4em\relax New York, NY, USA: Association for Computing Machinery, 2021, p. 307–319. [Online]. Available: \url{https://doi.org/10.1145/3433210.3453090}
\BIBentrySTDinterwordspacing

\bibitem{nasr2019comprehensive}
M.~Nasr, R.~Shokri, and A.~Houmansadr, ``Comprehensive privacy analysis of deep learning: Passive and active white-box inference attacks against centralized and federated learning,'' in \emph{2019 IEEE symposium on security and privacy (SP)}.\hskip 1em plus 0.5em minus 0.4em\relax IEEE, 2019, pp. 739--753.

\bibitem{10.1145/3133956.3134012}
\BIBentryALTinterwordspacing
B.~Hitaj, G.~Ateniese, and F.~Perez-Cruz, ``Deep models under the gan: Information leakage from collaborative deep learning,'' in \emph{Proceedings of the 2017 ACM SIGSAC Conference on Computer and Communications Security}, ser. CCS '17.\hskip 1em plus 0.5em minus 0.4em\relax New York, NY, USA: Association for Computing Machinery, 2017, p. 603–618. [Online]. Available: \url{https://doi.org/10.1145/3133956.3134012}
\BIBentrySTDinterwordspacing

\bibitem{diabetes}
D.~Dua and C.~Graff, ``Uci machine learning repository,'' \url{http://archive.ics.uci.edu/ml}, 2017, accessed: 2023-03-25.

\bibitem{adult}
\BIBentryALTinterwordspacing
------, ``{UCI} machine learning repository,'' 2017. [Online]. Available: \url{http://archive.ics.uci.edu/ml}
\BIBentrySTDinterwordspacing

\bibitem{Tran2017Online}
N.~N. Tran and J.~Lee, ``Online reviews as health data: Examining the association between availability of health care services and patient star ratings exemplified by the yelp academic dataset,'' \emph{JMIR Public Health Surveill}, vol.~3, no.~3, p. e43, 2017.

\bibitem{10.5555/3241094.3241142}
F.~Tram\`{e}r, F.~Zhang, A.~Juels, M.~K. Reiter, and T.~Ristenpart, ``Stealing machine learning models via prediction apis,'' in \emph{Proceedings of the 25th USENIX Conference on Security Symposium}, ser. SEC'16.\hskip 1em plus 0.5em minus 0.4em\relax USA: USENIX Association, 2016, p. 601–618.

\bibitem{gan-leak}
\BIBentryALTinterwordspacing
D.~Chen, N.~Yu, Y.~Zhang, and M.~Fritz, ``Gan-leaks: A taxonomy of membership inference attacks against generative models,'' in \emph{Proceedings of the 2020 ACM SIGSAC Conference on Computer and Communications Security}, ser. CCS '20.\hskip 1em plus 0.5em minus 0.4em\relax New York, NY, USA: Association for Computing Machinery, 2020, p. 343–362. [Online]. Available: \url{https://doi.org/10.1145/3372297.3417238}
\BIBentrySTDinterwordspacing

\end{thebibliography}

\begin{IEEEbiography}[{\includegraphics[width=1in,height=1.25in,clip,keepaspectratio]{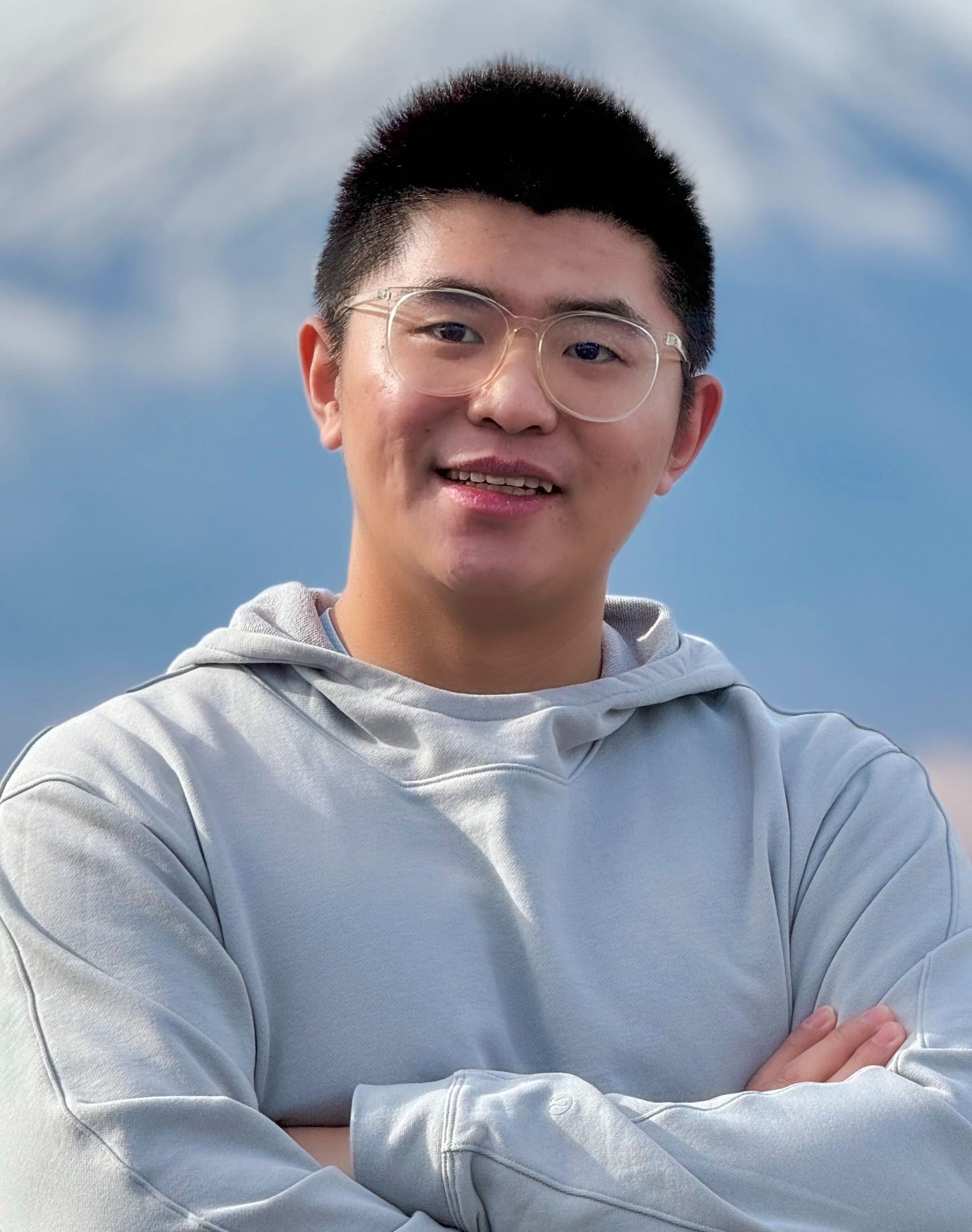}}]{Zhou Yang} is a Senior Research Egnineer and PhD candidate at Singapore Management University under the supervision of Prof. David Lo. He received his MSc degree in Software System Engineering from University College London and his B.Eng degree in Software Engineering from Yangzhou University. Zhou currently focuses on different properties of large language models of code, e.g., robustness, security, usability, etc. He has published papers in top-tier venues, including ICSE, FSE, ASE and ISSTA, and journals such as TSE, TOSEM, and EMSE. He won the ACM SIGSOFT Distinguished Paper Award and the ACM Student Research Competition Award. Zhou likes to walk on the streets and freeze memorable moments with his Fujifilm X100 camera. More information can be found at: https://yangzhou6666.github.io.
\end{IEEEbiography}

\begin{IEEEbiography}[{\includegraphics[width=1in,height=1.25in,clip,keepaspectratio]{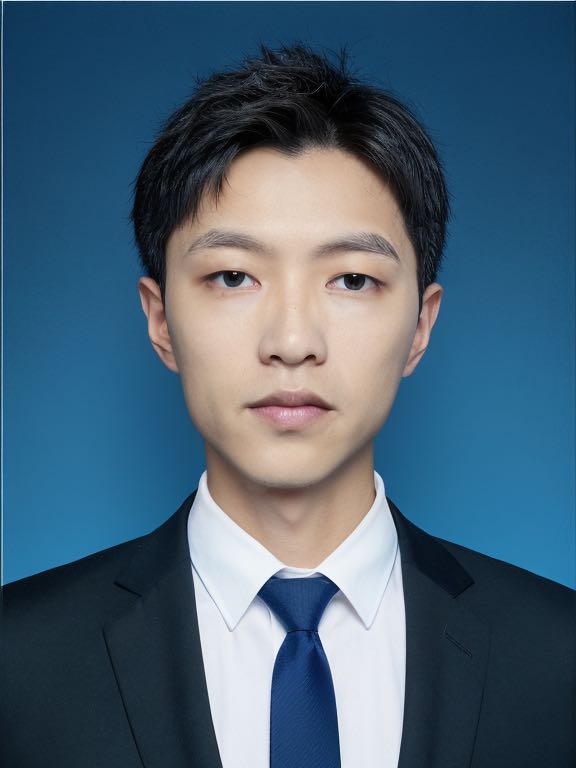}}]{Zhipeng Zhao} is a Master's student at the University of Copenhagen, with previous experience as a Research Assistant at Singapore Management University under the supervision of Prof. David Lo. His research focuses on the intersection of artificial intelligence (AI) and diverse fields (AI+X), particularly exploring how AI can drive innovation, enhance problem-solving capabilities, and tackle complex challenges across multiple disciplines. His work has been featured in top-tier conferences, including ICSE and RecSys. More information: https://oran-ac.github.io.
\end{IEEEbiography}

\begin{IEEEbiography}[{\includegraphics[width=1in,height=1.25in,clip,keepaspectratio]{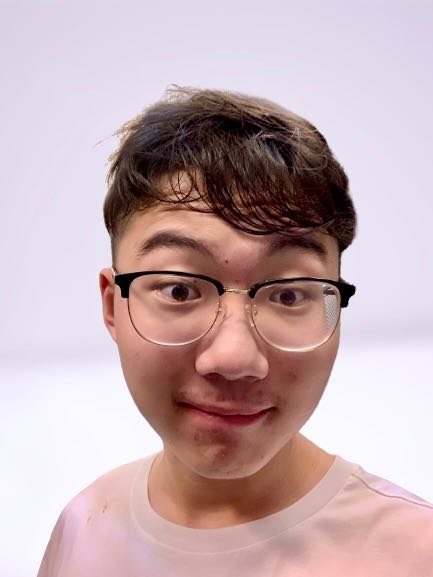}}]{Chenyu Wang} is a PhD candidate and a Research Engineer at the School of Computing and Information Systems (SCIS), Singapore Management University (SMU). His research interests lie primarily in the intersection of Software Engineering (SE) and Artificial Intelligence (Al). He specifically focuses on two areas: (1) the quality assurance of AI-based systems, and (2) backdoor attacks on large language models for code (LLM4Code). His work has been published in high-quality SE conferences \& journals such as ICSE, TSE, MSR, etc. Contact him: chenyuwang@smu.edu.sg. More info: https://github.com/JamesNolan17.
\end{IEEEbiography}

\begin{IEEEbiography}[{\includegraphics[width=1in,height=1.25in,clip,keepaspectratio]{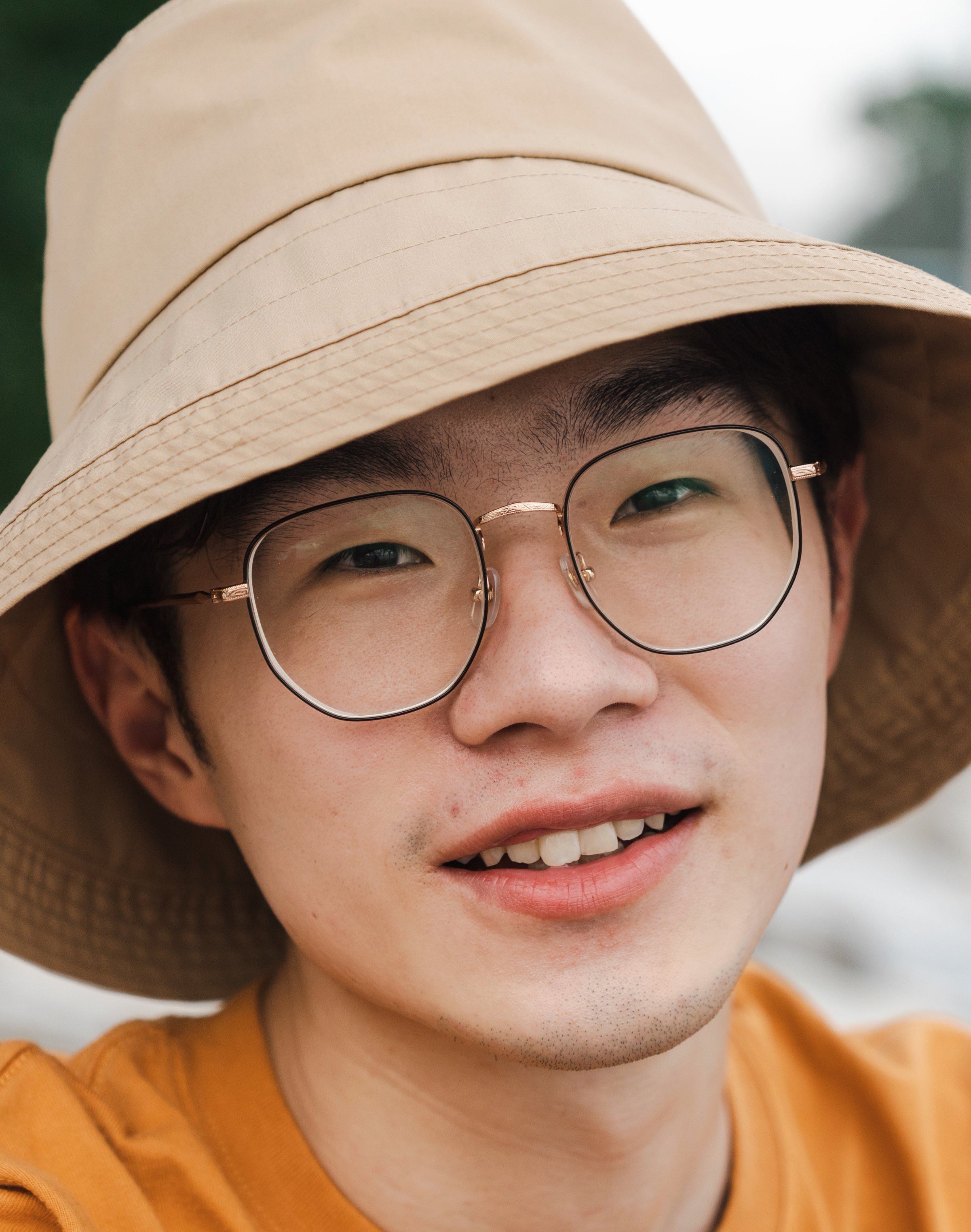}}]{Jieke Shi} s a PhD candidate and a Research Engineer at the School of Computing and Information Systems (SCIS), Singapore Management University (SMU). His research interests lie primarily in the intersection of Software Engineering (SE) and Artificial Intelligence (AI). Particularly, he focuses on (1) quality assurance of AI-enabled systems from an SE perspective, and (2) efficiency improvement of code models for real-world deployment. His work has been published in high-quality SE conferences such as ICSE, ASE, MSR, etc. He has won/been nominated for several research paper awards. Contact him: jiekeshi@smu.edu.sg. More info: https://jiekeshi.github.io.
\end{IEEEbiography}

\begin{IEEEbiography}[{\includegraphics[width=1in,height=1.25in,clip,keepaspectratio]{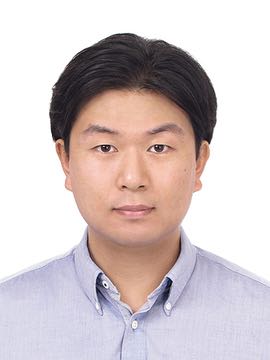}}]{Dongsun Kim}
  is an associate professor at Korea University.
  He has received his Ph.D. degree in Computer Science and Engineering from Sogang University, Korea. His career includes several academic and industrial experiences in Hong Kong, Luxembourg, and Korea.
  He has published several research papers and participated in several research projects relevant to automated software engineering. 
  In particular, he has pioneered a new line of research on pattern-based program repair.
  His recent achievements have focused on proactive debugging, automated fix pattern mining, deep code representation for mining fix patterns, program repair driven by bug reports, fault localization impact on program repair, and specific topics for program repair.
\end{IEEEbiography}

\begin{IEEEbiography}[{\includegraphics[width=1in,height=1.25in,clip,keepaspectratio]{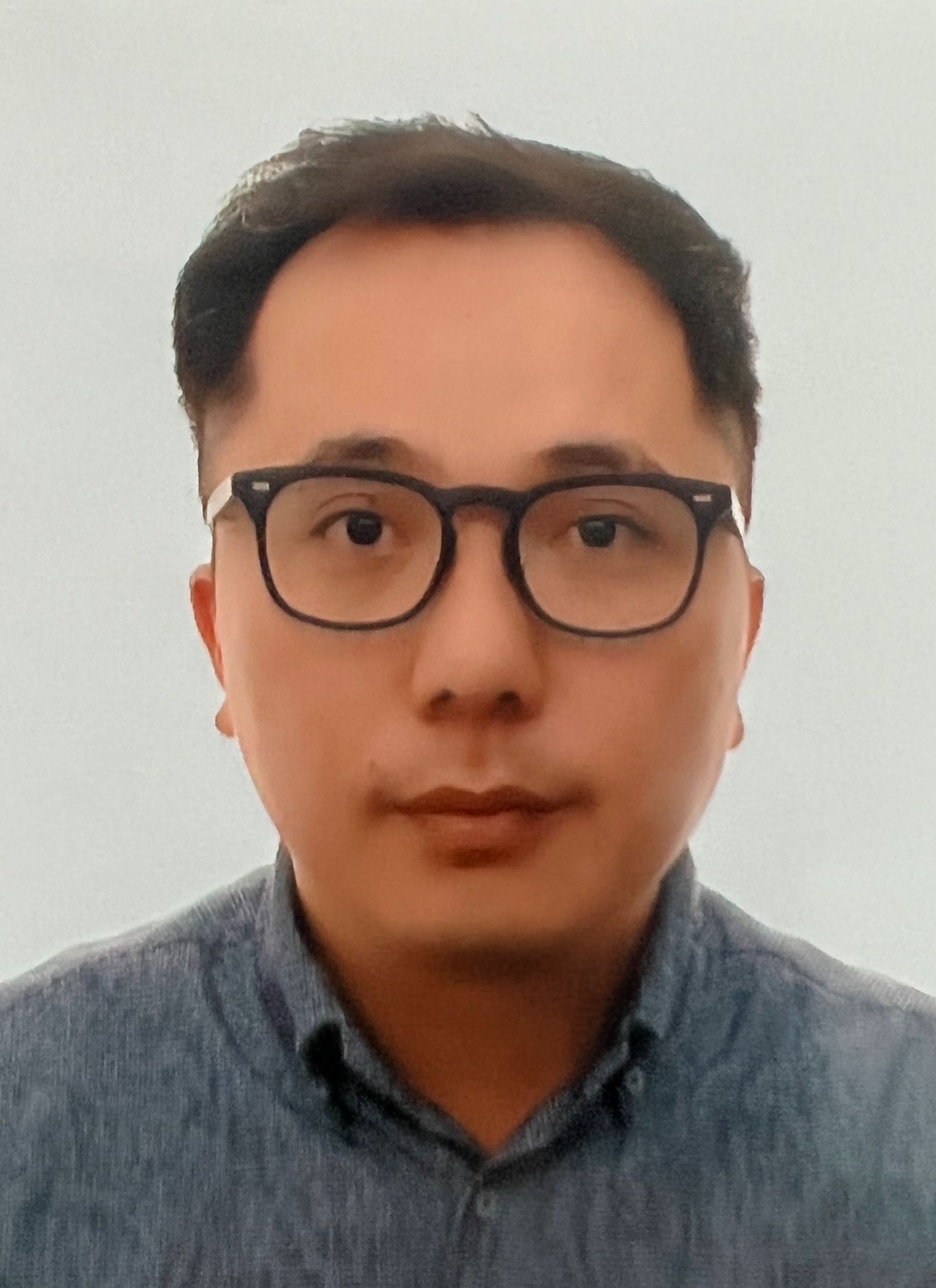}}]{DongGyun Han} is a lecturer (assistant professor) at the Department of Computer Science at Royal Holloway, University of London. He is mainly working on software engineering research. His main research interests are Empirical Study, AI for Software Engineering (AI4SE), Software Engineering for AI (SE4AI), and Code Review. He was a research scientist (postdoc) at SOftware Analytics Research (SOAR) group and Secure Mobile Centre at Singapore Management University (SMU). Before joining SMU, he was a software development engineer at Amazon Web Services. He completed his PhD at the University College London (UCL). He has worked for KAIST Institute for IT Convergence as a researcher after getting his MPhil. at Hong Kong University of Science and Technology (HKUST).
\end{IEEEbiography}

\begin{IEEEbiography}[{\includegraphics[width=1in,height=1.25in,clip,keepaspectratio]{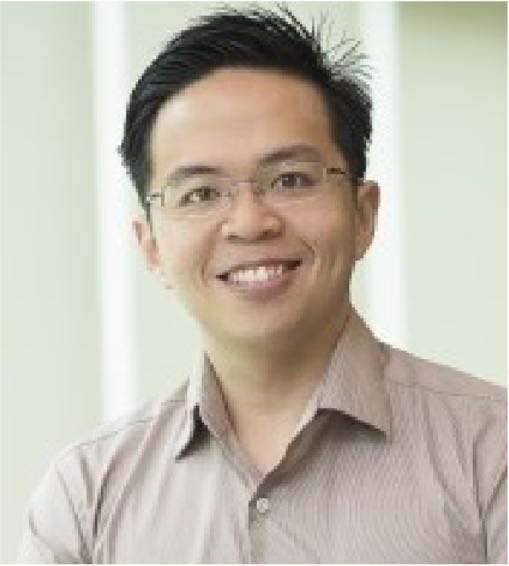}}]{David Lo} is the OUB Chair Professor and the Director of the Information Systems and Technology Cluster, School of Computing and Information Systems, Singapore Management University. His research interest is in the intersection of software engineering, cybersecurity and data science, encompassing socio-technical aspects and analysis of different kinds of software artefacts, with the goal of improving software quality and security and developer productivity. He has won more than 20 international research and service awards including more than 10 ACM SIGSOFT / IEEE TCSE Distinguished Paper Awards and the 2021 IEEE TCSE Distinguished Service Award. He has served in more than 40 organizing committees, including serving as a General/Program Co-Chair of ICSE 2025, ESEC/FSE 2024, MSR 2022, ASE 2020, SANER 2019, and ICSME 2018. He has also served on the Editorial Boards of a number of journals including IEEE Transactions on Software Engineering, Empirical Software Engineering , and IEEE Transactions on Reliability. He is an ACM Fellow (since 2023), IEEE Fellow (since 2022), and ASE Fellow (since 2021).
\end{IEEEbiography}

\IEEEdisplaynontitleabstractindextext

\end{document}